\pdfoutput=1

\documentclass[11pt]{article}

\usepackage[preprint]{acl}

\usepackage{times}
\usepackage{latexsym}

\usepackage[T1]{fontenc}

\usepackage[utf8]{inputenc}

\usepackage{microtype}

\usepackage{inconsolata}

\usepackage{graphicx}

\usepackage{makecell}
\usepackage{tcolorbox}
\usepackage{subcaption}
\usepackage{wrapfig}

\title{3MDBench: Medical Multimodal Multi-agent Dialogue Benchmark}

\author{
  \textbf{Ivan Sviridov\textsuperscript{1*}},
  \textbf{Amina Miftakhova\textsuperscript{1*}},
  \textbf{Artemiy Tereshchenko\textsuperscript{1}},
  \textbf{Galina Zubkova\textsuperscript{1}},
  \\
  \textbf{Pavel Blinov\textsuperscript{1}},
  \textbf{Andrey Savchenko\textsuperscript{1,2,3}}
\\
\\
 \textsuperscript{1}Sber AI Lab, Moscow, Russia,
 \textsuperscript{2}HSE University, Moscow, Russia \\
 \textsuperscript{3}ISP RAS Research Center for Trusted Artificial Intelligence, Moscow, Russia 
 \\
\small{
\textbf{Correspondence:} \href{wchhiaarid@gmail.com}{wchhiaarid@gmail.com}, \href{mailto:noteisenheim@gmail.com}{noteisenheim@gmail.com},
\href{mailto:blinoff.pavel@gmail.com}{blinoff.pavel@gmail.com},
\href{mailto:avsavchenko@hse.ru}{avsavchenko@hse.ru}
 }
}

\begin{document}
\maketitle
\begin{abstract}

    Though Large Vision-Language Models (LVLMs) are being actively explored in medicine, their ability to conduct complex real-world telemedicine consultations combining accurate diagnosis with professional dialogue remains underexplored. This paper presents \textbf{3MDBench} (\textbf{M}edical \textbf{M}ultimodal \textbf{M}ulti-agent \textbf{D}ialogue \textbf{Bench}mark), an open-source framework for simulating and evaluating LVLM-driven telemedical consultations. 3MDBench simulates patient variability through temperament-based Patient Agent and evaluates diagnostic accuracy and dialogue quality via Assessor Agent. It includes 2996 cases across 34 diagnoses from real-world telemedicine interactions, combining textual and image-based data. The experimental study compares diagnostic strategies for widely used open and closed-source LVLMs. We demonstrate that multimodal dialogue with internal reasoning improves F1 score by 6.5\% over non-dialogue settings, highlighting the importance of context-aware, information-seeking questioning. Moreover, injecting predictions from a diagnostic convolutional neural network into the LVLM's context boosts F1 by up to 20\%. Source code is available at \url{https://github.com/univanxx/3mdbench}.

\end{abstract}
\def\thefootnote{*}\footnotetext{These authors contributed equally to this work.}\def\thefootnote{\arabic{footnote}}

\section{Introduction}
Telemedicine expands healthcare access and efficiency by enabling real-time consultations and early diagnosis~\cite{Stoltzfus2023}. In these consultations, effective communication is essential for diagnostic accuracy and treatment adherence~\cite{info:doi/10.2196/16965, 10.3389/fpubh.2020.00410}. Large Language Models (LLMs) and Vision-Language Models (LVLMs) further enhance telehealth via real-time analysis~\cite{IJLSRA-2024-0061}, chronic care management~\cite{56671b465928b43c402555fc6e5412bf3d78}, and decision support~\cite{healthcare-13-00324}, including symptom assessment, test interpretation, and patient interaction~\cite{Blinov_2024,Lu2024-sf, 1-s2.0-S0747563224002875-main, kumichev2024medsyn}.

\begin{figure*}[t]
  \centering
  \includegraphics[width=\linewidth]{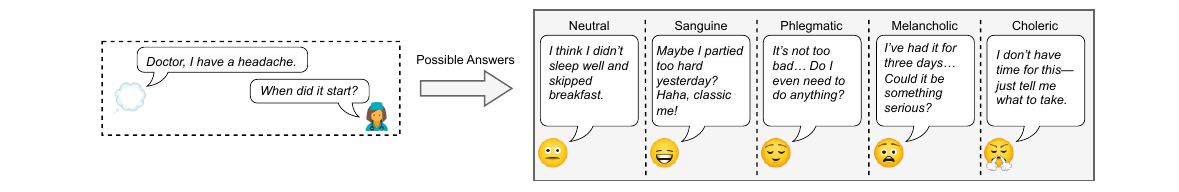}
  \caption{The patient's response during the dialogue with the doctor depends on the temperament.}
  \label{fig:teaser}
\end{figure*}

Despite recent advances in LLMs' application in telemedicine, effective doctor-patient dialogue remains essential. Diagnostic quality depends on how patients articulate symptoms, shaped by emotional, cognitive, and systemic factors~\cite{amelung, singh_sittig}. Miscommunication, fear, and low health literacy delay diagnosis~\cite{heyhoe, nguyen}, while temperament influences responsiveness, expressiveness, and trust~\cite{GraedonGraedon+2014+49+51, Meyeretal+diag+accuracy+confidence}. However, existing LLM benchmarks for assessing telemedicine quality offer limited realism by restricting models to multiple-choice tasks~\cite{jin2020diseasedoespatienthave, kim2024medexqamedicalquestionanswering}, imposing factual and non-reactive dialogue templates with artificial interruptions that prevent full consultations~\cite{li2024mediqquestionaskingllmsbenchmark, Johri2023.09.12.23295399}, and omitting image modalities representing patient symptoms~\cite{zhu2025askpatientspatienceenabling}.


To address these problems, we introduce \textbf{3MDBench} (\textbf{M}edical \textbf{M}ultimodal \textbf{M}ulti-agent \textbf{Dialogue} Benchmark) to evaluate LVLM-based consultations in dynamic, realistic scenarios. We use classical temperament theory, which categorizes individuals into four types (sanguine, choleric, melancholic, and phlegmatic)~\cite{steiner1985four}, to model the personal characteristics that might affect engagement and medical alliance~\cite{Paap2022-za, neurosci4030017}. Hence, our 3MDBench features a \textbf{Patient Agent}, simulating one of four temperament types shown in Fig.~\ref{fig:teaser}, a \textbf{Doctor Agent}, LVLM to conduct the telemedicine consultation and serving as the benchmarked component, and an \textbf{Assessor Agent}, evaluating diagnostic accuracy and communication quality. We select models that best match defined metrics and human annotations to construct these agents. Built on 34 diagnoses obtained from real-world telemedicine consultations using medical image datasets enriched with textual information, 3MDBench supports both textual and image modalities. We benchmark commercial and open-source LVLMs in different dialogue and non-dialogue settings. Our results show that information-seeking dialogue strategies using medical reasoning and image modality increase the F1 score to 6.5\%, highlighting the importance of adaptive, context-aware interaction. Finally, our novel approach, incorporating top-3 predictions from a ConvNet trained on diagnosis into the LVLM's context, boosts the F1 score up to 20\%.

In this work:
\begin{enumerate}
    
    \item We propose \textbf{3MDBench}, an open-source benchmark for evaluating medical dialogue systems, with an Assessor Agent measuring diagnostic accuracy and communication quality, and a Patient Agent simulating temperament-based personality-driven doctor-patient conversations.  
    
    \item Using our standardized framework for assessing AI-driven medical consultation quality, we extensively compare open-source and state-of-the-art LVLMs.

    \item We analyze multiple dialogue strategies for a Doctor Agent: with and without image modality, implementing rationale generation, and using external cues. We demonstrate the importance of image modality and conducting information-seeking conversations with internal reasoning. Moreover, we show that the diagnostic F1 score of the Doctor Agent improves up to 20\% by incorporating top-3 predictions from a specially trained diagnostic convolutional neural network into the LVLM.
\end{enumerate}
\begin{table*}[ht]
\caption{Comparison of 3MDBench with existing medical benchmarks and datasets.  The columns are: \textbf{T} (Type: Dataset (DS) or Benchmark (BM)), \textbf{TDT} (Text Data Type: Question-Answer pairs (QA) or Dialogues (D)), \textbf{N} (Name of Dataset/Benchmark), \textbf{M} (Modality: Text-only (T) or Multimodal (M)), \textbf{S} (Size of test part of a Benchmark of full size of a Dataset), \textbf{D} (Dialogues present), \textbf{A} (Multi-Agent approach used), \textbf{P} (Personality modeling used), \textbf{CQ} (Consultation and communication qualities tested), \textbf{F} (Full-fledged consultation simulated until both agents naturally conclude the dialogue), and \textbf{L} (Language of data).}\label{tab:translated_table}
\resizebox{\textwidth}{!}{
\begin{tabular}{ccccccccccl}
\hline
\textbf{T} & \textbf{TDT} & \textbf{N} & \textbf{M} & \textbf{S} & \textbf{D} & \textbf{A} & \textbf{P} & \textbf{CQ} & \textbf{F} & \textbf{L} \\ 
\hline
DS & D & MedDialog-EN~\cite{Zeng_2020_MedDialog} & T & 300K & + & - & - & - & - & EN \\ 
DS & D & MedDialog-CN~\cite{Zeng_2020_MedDialog} & T & 1100K & + & - & - & - & - & CN \\
DS & D & MedDG~\cite{Liu_2022_MedDG} & T & 18K & + & - & - & - & - & CN \\
DS & D & CMtMedQA~\cite{Yang_2023_CMtMedQA} & T & 70K & + & - & - & - & - & CN \\ 
DS & D & Icliniq-10K~\cite{li2023ChatDoctor}  & T & 10K & + & - & - & - & - & EN \\ 
DS & D / QA & BianQueCorpus~\cite{Yirong_2023_BianQue} & T & 2437K & + & - & - & - & - & CH \\ 
DS & D / QA & HealthCareMagic-100k~\cite{Li_2023_HealthCareMagic} & T & 100K & + & - & - & - & - & EN \\  
DS & D / QA & Psych8k~\cite{Yuan_2025_Psych8k} & T & 8K & + & - & - & - & - & EN \\ 
DS & D & IMCS-21~\cite{Chen_2022_IMCS} & T & 811 & + & + & - & - & - & CN \\ 
DS & D & NoteChat~\cite{Wang_2024NoteChat} & T & 30K & + & + & - & - & - & EN \\
DS & D & MTMedDialog~\cite{feng2025doctoragentrlmultiagentcollaborativereinforcement} & T & 10.1K & + & + & - & - & - & EN \\
\hline
BM & QA & Cholec80-VQA ~\cite{Twinanda_2016_Cholec80} & M & 9K & - & - & - & - & - & EN \\ 
BM & QA & VQA-RAD~\cite{Lau_2018_VQA-RAD} & M &3.5K & - & - & - & - & - & EN \\ 
BM & QA & PathVQA ~\cite{He_2020_PathVQA} & M & 6K & - & - & - & - & - & EN \\ 
BM & QA & SLAKE~\cite{Liu_2021_SLAKE} & M & 2K & - & - & - & - & - & EN \\
BM & QA & RadBench ~\cite{harrison_ai_radbench} & M & 137K & - & - & - & - & - & EN \\ 
BM & QA & MMMU (H \& M)~\cite{Yue_2024_MMMU} & M & 11.5K & - & - & - & - & - & EN \\  
BM & QA & OmniMedVQA~\cite{Hu_2024_OmniMedVQA} & M & 128K & - & - & - & - & - & EN \\ 
BM & QA & GMAI-MMBench~\cite{Chen_2024_GMAI} & M & 26K & - & - & - & - & - & EN \\
BM & QA & Medical-Diff-VQA ~\cite{Hu_2025_MedicalDiff} & M & 70K & - & - & - & - & - & EN \\ 
BM & D & MediQ~\cite{li2024mediqquestionaskingllmsbenchmark} & T & 1.2K & + & + & - & - & - & EN \\
BM & D & AgentClinic~\cite{schmidgall2024agentclinic} & M & 457 & + & + & - & - & - & EN \\
BM & D & MedAgentSim~\cite{almansoori2025selfevolvingmultiagentsimulationsrealistic} & M & 637 & + & + & - & - & - & EN \\
BM & D & AI Hospital~\cite{fan2024aihospitalbenchmarkinglarge} & M & 506 & + & + & + & + & - & CN \\
BM & D & Dr.APP~\cite{zhu2025askpatientspatienceenabling} & T & 1.5K & + & + & + & + & - & EN \\
\hline
BM & D & 3MDBench (Ours) & M & 3K & + & + & + & + & + &EN \\ 
\hline
\end{tabular}
}
\end{table*}

\section{Related Work}

Benchmarks for medical LLMs have focused primarily on factual knowledge, evaluating performance on exams, QA tasks, and case-based reasoning~\cite{jin2020diseasedoespatienthave, kim2024medexqamedicalquestionanswering, pmlr-v174-pal22a, jin2019pubmedqa, Singhal2023} to assess domain understanding and consistency, but overlook interactive and contextual aspects of diagnosis. Recent work highlights the need for dialogue-based evaluation, where models must elicit, interpret, and reason over patient-reported symptoms~\cite{10.1001/jamanetworkopen.2024.40969, manes2024KQA, li2023ChatDoctor, han2023MedAlpaca}. However, existing benchmarks often rely on scripted interactions or fixed-response patient agents, limiting their reflection of real-world consultations~\cite{shi-etal-2024-medical}.

Multi-agent systems offer a more dynamic alternative, as they simulate collaborative diagnostic workflows, enabling LLMs to interact, reason, and refine decisions over multiple turns~\cite{qiu2024llm}. Recent efforts have extended this paradigm to simulate patient interactions in telemedicine consultations~\cite{kim2024mdagents, mehandru2024evaluating, li2024mediqquestionaskingllmsbenchmark, schmidgall2024agentclinic, fan2024aihospitalbenchmarkinglarge, zhu2025askpatientspatienceenabling, almansoori2025selfevolvingmultiagentsimulationsrealistic}.

Although these benchmarks represent important progress, none of them simultaneously addresses the following critical limitations:

\begin{itemize}
  \item Simulating patients as static and personality-free, reducing interactions to factual inputs and overlooking how traits like emotion, communication style, or temperament influence diagnostic accuracy~\cite{amelung}. 
  \item Focusing solely on diagnostic and recommendation accuracy, overlooking comprehensive consultation and communication quality assessment based on full-fledged, realistic dialogue—an essential dimension that directly influences patient trust and further diagnostic outcomes~\cite{Ha2010-ov}.
  \item Excluding image modality from the diagnostic process, despite its significant role in real-world diagnostic decision-making~\cite{AGBAREIA2025184}.
\end{itemize}

To overcome these limitations, we propose \textbf{3MDBench} that simulates and evaluates telemedicine consultation with a temperament-driven Patient Agent and an Assessor Agent for accuracy and communication quality. Compared to existing benchmarks (Table~\ref{tab:translated_table}), we capture the variability and complexity of real-world clinical interactions, enabling richer, more patient-aligned evaluation of medical dialogue systems.
\begin{figure*}[t]
  \includegraphics[width=0.48\linewidth]{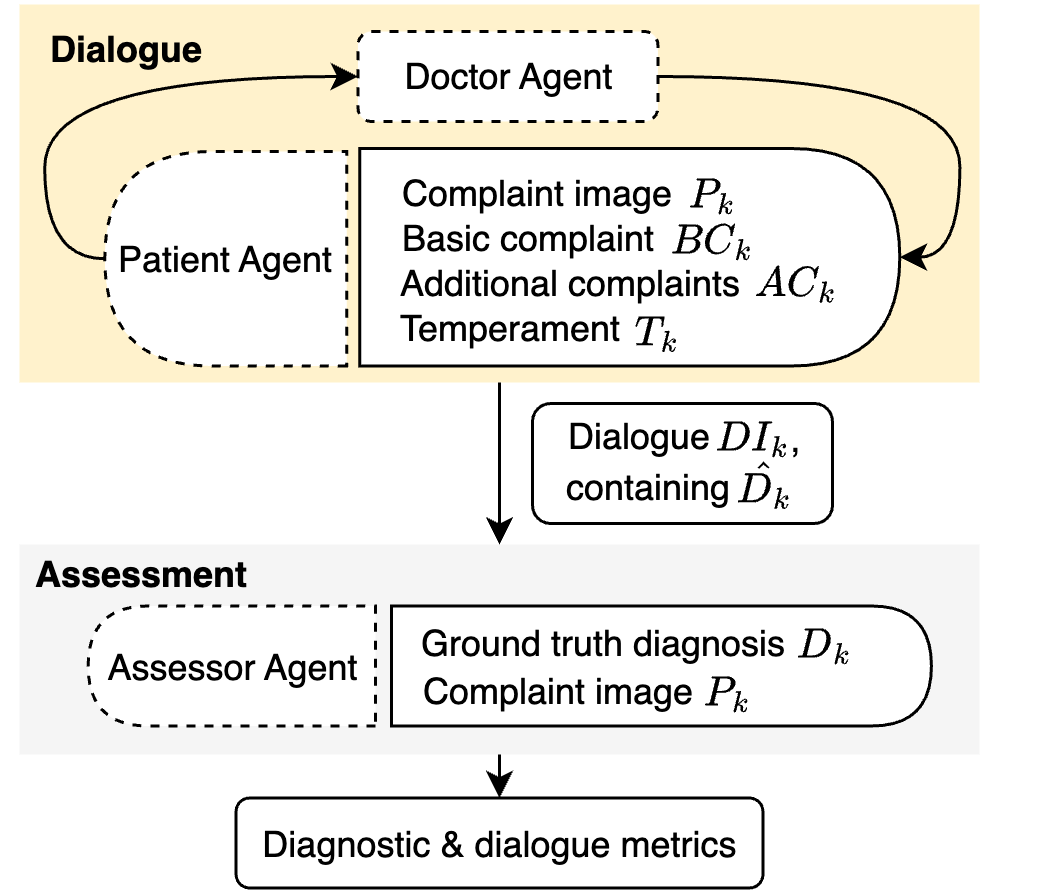} \hfill
  \includegraphics[width=0.48\linewidth]{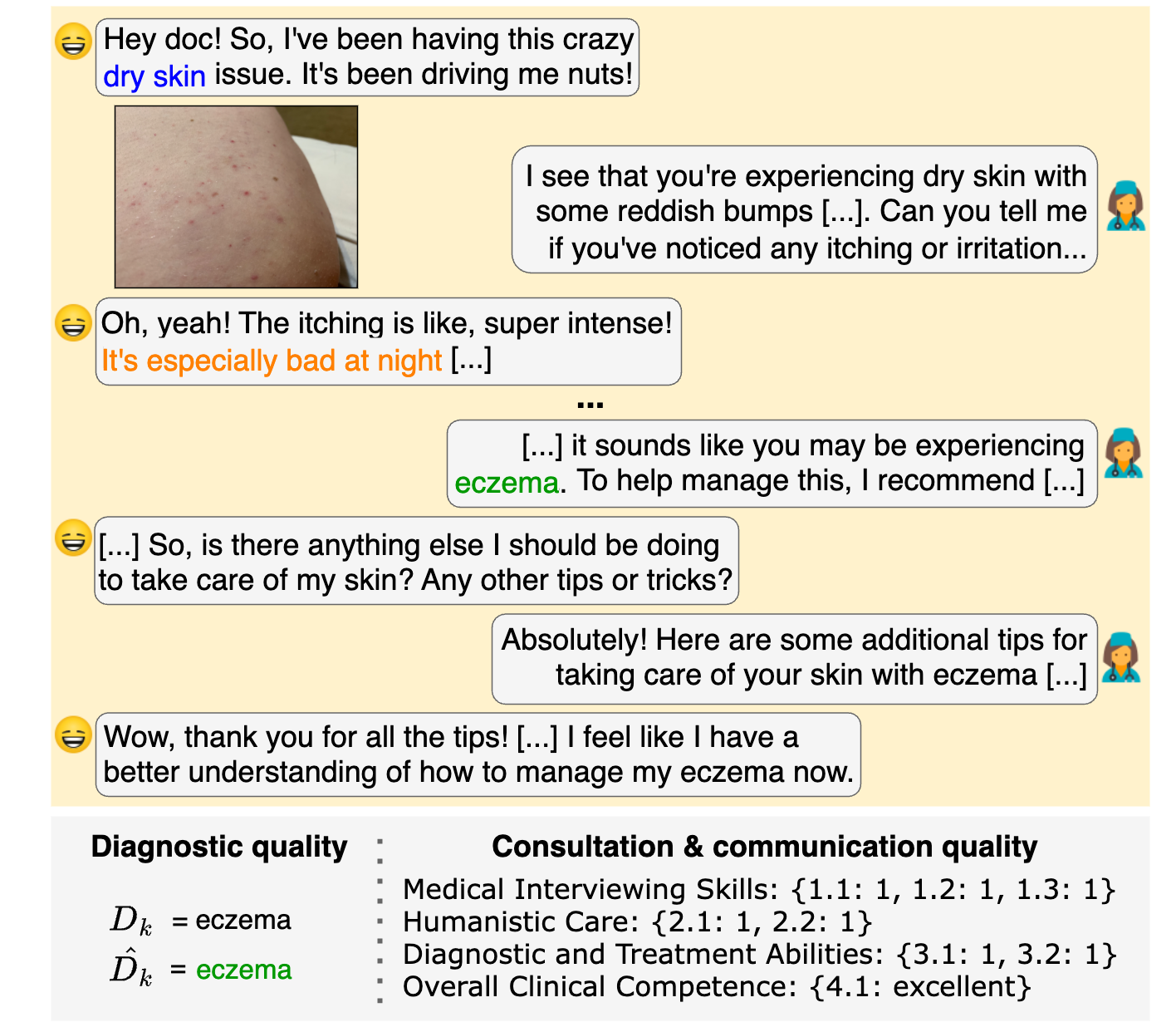}
  \caption {Agents' interaction pipeline in 3MDBench (left) and an example instance (right). The dialogue \( DI_k \) begins with the Patient Agent of sanguine temperament \(T_k\), whose first reply contains the complaint image \(P_k\) and the basic complaint \( BC_k \) highlighted in blue; an additional complaint from the list \(AC_k\) appears later in orange, and the final diagnosis \(\hat{D}_k\)—identified and validated by the Assessor Agent—is shown in green. The Assessor Agent, based on \(DI_k\) and \(\hat{D}_k\), further provides a structured evaluation of diagnostic performance as well as consultation and communication quality.}
  \label{fig:scheme}
\end{figure*}

\section{Proposed 3MDBench}
\subsection{Data Collection}\label{real_data_check}

\paragraph{Diagnoses.}

To ensure clinical relevance, we analyzed 611K anonymized visits from a large Eastern European provider from May to October 2024, selecting the top 80\% most frequent diagnoses. We examined 180 million outpatient records from the same city through 2022 to validate cross-setting consistency. All diagnoses, originally in ICD-10~\cite{10665-42980}, were standardized using a physician-curated dictionary. The final set comprises 34 diagnoses across five medical domains as Figure~\ref{fig:real_distr} of Appendix~\ref{app:3mdbench_table} shows.

\paragraph{Image Data.}

We constructed 3MDBench from 6 open-source datasets, primarily from Kaggle \footnote{\url{https://github.com/Priyanshu9898/Oral-Disease-Classification}} \footnote{\url{https://www.kaggle.com/datasets/anindamohanta/different-phases-of-tonsilitis}} \footnote{\url{https://www.kaggle.com/datasets/nikhilgurav21/nail-disease-detection-dataset}} \footnote{\url{https://www.kaggle.com/datasets/alisofiya/conjunctivitis}}, as well as ISIC Archive images~\cite{CASSIDY2022102305}, Google SCIN~\cite{10.1001/jamanetworkopen.2024.46615}, and Fitzpatrick17k~\cite{groh2021evaluating}, with supplemental data using Bing Image Search\footnote{\url{https://www.microsoft.com/en-us/bing/apis/bing-image-search-api}}~\cite{ghosh2023clipsyntelclipllmsynergy}. We adjusted disease prevalence to match the distribution from a primary Eastern European telemedicine provider described above to align with real-world telemedicine diagnosis distribution. To ensure sufficient evaluation data and mitigate class imbalance, we set a minimum threshold of 64 images per condition, an empirically determined lower bound based on the maximum number of images available for certain classes across all sources and the Internet. Collected images were filtered through automated quality checks (e.g., size, blur, watermarks) and manual review by one of the coauthors with medical expertise. The final benchmark contains 2,996 images, with class distribution detailed in Figure~\ref{fig:data_distr} in Appendix~\ref{app:3mdbench_table}, plus private training and validation sets with 2,396 and 596 images, respectively.

\paragraph{Enriching Images with Textual Data.}  

To enrich the Patient Agent's input and support more natural telemedicine dialogues, we generated concise, image-associated descriptions for all 2,996 cases. First, using GPT-4o-mini~\cite{openai2024gpt4ocard}, selected for its high medical accuracy and relatively low cost~\cite{Li2023-ky, 10.1093/jamiaopen/ooae114,Abrar2025llmevalonconsumerhealth},  we generated one basic symptom from a human perspective for each of the 34 diagnoses. Then, we expanded each corresponding image for all cases, generating additional structured complaints describing affected areas, duration, intensity, and relevant patient history. This enriched textual input, with a generation prompt in Appendix~\ref{app:text_gen} and examples in Appendix~\ref{app:examples}, enhances the visual data and provides context for more informative interactions with the Doctor Agent.

\subsection{Task Definition}

3MDBench simulates realistic medical consultations via multi-turn dialogues between an evaluated \textbf{Doctor Agent} and a \textbf{Patient Agent} with further evaluation of an \textbf{Assessor Agent}, as Figure~\ref{fig:scheme} shows. Each scenario includes a ground truth diagnosis \( D_k \), a complaint image \( P_k \), and symptoms split into a \textbf{basic complaint} \( BC_k \) (shared initially) and \textbf{additional complaints} \( AC_k \) (revealed during dialogue). The Patient Agent, shaped by a predefined temperament \( T_k \), starts with access to \( BC_k \), \( P_k \), and \( AC_k \), and interacts using prompts from Appendix~\ref{app:3mdbench}, aiming to obtain a diagnosis and medical recommendations from the Doctor Agent.


The Doctor Agent receives an initial Patient Agent query containing only \( BC_k \) and \( P_k \), aims to uncover \( AC_k \) through dialogue with prompts from Appendix~\ref{app:doctors_agents}, and outputs diagnosis, treatment plan, and recommendations. Dialogues are capped at 28 utterances, matching the average length from real dialogues described in Subsection~\ref{real_data_check}, and are marked incomplete if unresolved within this limit.

The Assessor Agent evaluates complete dialogue \( DI_k \) by comparing extracted diagnosis \( \hat{D}_k \) with \( D_k \), and assessing diagnostic reasoning, communication, and clinical accuracy, based on the prompt in Appendix~\ref{app:3mdbench}.

\begin{table*}
  \caption{Criteria for doctor model assessment}
  \label{tab:doctor_criteria}
  \resizebox{\textwidth}{!}{
  \footnotesize
  \begin{tabular}{p{2.5cm}p{13.5cm}}
    \hline
    \textbf{Primary Item} & \textbf{Secondary Item} \\
    \hline
    Medical Interviewing Skills & 1.1: Does the doctor enquire about a patient's medical history, such as previous diseases, medications, and surgeries? \\
    & 1.2: Does the doctor enquire about the current symptoms, possible causes, and attempted treatments? \\
    & 1.3: Does the doctor explain the basis of the provided conclusion to the patient? \\
    \hline
    Humanistic Care & 2.1: Does the doctor communicate with respect, empathy, and politeness, providing appropriate guidance and avoiding unnecessary extensions? \\
    & 2.2: Does the doctor respect the individual wishes of the patient? \\
    \hline
    Diagnostic and & 3.1: Does the doctor provide an accurate diagnostic plan for the supposed diagnosis? \\
    Treatment Abilities & 3.2: Does the doctor accurately provide a treatment plan for the supposed diagnosis? \\
    \hline
    Overall Clinical Competence & 4.1: Which level of clinical competence does the doctor demonstrate during the consultation? (Unsatisfactory, satisfactory, or excellent). \\
    \hline
  \end{tabular}
  }
\end{table*}

\subsection{Patient Agent}\label{subsec:patient}

The quality of the benchmark depends on the performance of the Patient Agent. This agent, lacking access to the ground-truth diagnosis but aware of its symptoms, engages in text-based dialogue and concludes once the doctor provides a diagnosis, recommendations, and answers all questions.

To ensure that candidate models cover different families and provide various strategies, we selected for our Llama-3-8B-Instruct and Llama-3.1-8B from the Llama family~\cite{grattafiori2024llama3herdmodels}, Qwen2.5-7B and Qwen2.5-14B from the Qwen family~\cite{yang2024qwen2technicalreport}, Falcon-7B~\cite {almazrouei2023falconseriesopenlanguage}, and GPT-4o-mini.

Patient Agent must strictly follow system prompts provided in Appendix~\ref{app:pat_select}, respond relevantly to doctor queries, and remain truthful. Suitability to these requirements was assessed using three metrics. \textbf{Instruction following} (1–5 scale) measures prompt adherence, evaluated by GPT-4o-mini due to its strong performance in medical evaluation~\cite{info:doi/10.2196/56500}. \textbf{Relevance} is a binary metric that assesses whether each patient's response aligns with the doctor’s utterance, averaged over the dialogue, evaluated by GPT-4o-mini. \textbf{Factuality} measures how often utterances reference prompted symptoms, computed via NV-Embed-v2~\cite{lee2024nv} embeddings with a cosine similarity threshold of 0.8~\cite{li2024mediqquestionaskingllmsbenchmark}.


\subsection{Assessor Agent}\label{part:ass_agent}
 
Assessor Agent is responsible for \textbf{evaluating doctor agents} in generated dialogues and \textbf{extracting final diagnoses} from the doctor's conclusions. To assess the clinical competence, we adapted our evaluation criteria from the Mini-Clinical Evaluation Exercise (Mini-CEX)~\cite{shi-llm-mini-cex}, a standard in medical education, where patients evaluate medical consultations via structured questions. We simplified its 24 criteria by removing irrelevant items (e.g., autonomy, bias) and merging redundant ones, resulting in 8 core criteria (Table~\ref{tab:doctor_criteria}).

To select the best model, we measured alignment with human annotations on a diverse validation subset of 3MDBench, balanced across four patient temperaments, multiple doctor models, and 34 diagnoses. Five human annotators rated dialogues and extracted diagnoses using the exact instructions as the Assessor Agent (Appendix~\ref{bench-prompt-assessor-diag}). Inter-annotator agreement, measured using Cohen's Kappa and described in Appendix~\ref{sec:agreement}, yielded an average score of 0.49, indicating moderate agreement according to established interpretation scales~\cite{Artstein2008-agreement}. Given the complexity and subjectivity of clinical assessment leading to variability in human judgments, this level of agreement is standard across medical domains~\cite{Haas1996-gi, Verma2016-nu, cancers13215378}.


Next, we collect annotations and extract diagnoses from LVLM-based assessor-candidates using two prompts from Appendix~\ref{bench-prompt-assessor}. The candidate assessor models are Qwen2-VL-72B-Instruct~\cite{Qwen-VL}, GPT-4o-mini, Llava-OneVision-Qwen2-72b-ov-chat-hf~\cite{li2024llavaonevisioneasyvisualtask}, and DeepSeek-VL~\cite{lu2024deepseekvl}. Each model received the same input as human annotators in clinical competence evaluation: the dialogue, image, and ground truth diagnosis. 
\subsection{Evaluated Doctor Agents}
\label{sec:doctor}
The primary goal of 3MDBench is to evaluate the diagnostic capabilities of LVLMs in a simulated telemedicine setting. Specifically, the benchmark assesses a model's ability to integrate visual and textual modalities to emulate the role of a doctor during a consultation. At the start of each appointment, the doctor model receives a supporting medical image and is expected to engage in an information-seeking dialogue with the patient. The model aims to arrive at an accurate diagnosis informed by the image and the dialogue.

\begin{table*}[h!]
  \caption{Comparison of candidate patient models assessed in the diagnostic conversation using GPT-4o-mini based on the three important aspects. Then, the models are ranked based on each aspect, and the mean rank is calculated.}
  \label{tab:patient_models}
  \centering
  \resizebox{0.95\textwidth}{!}{
  \begin{tabular}{p{2cm}cccccc}
    \hline
    Model Name & Llama-3-8b & Llama-3.1-8b & Qwen2.5-7B & Qwen2.5-14B & Falcon-7B & GPT-4o-mini \\
    \hline
    Instruction following & 4.72 & \textbf{4.74} & 4.71 & 4.59 & 4.37 & 4.38 \\
    Relevance & 0.65 & 0.59 & 0.84 & 0.76 & \textbf{0.90} & 0.82 \\
    Factuality & 0.79 & 0.77 & 0.67 & 0.78 & 0.59 & \textbf{0.98} \\
    \hline
    Mean Rank & \textbf{3.00} & 3.67 & 3.33 & 3.67 & 4.33 & \textbf{3.00} \\
    \hline
  \end{tabular}
  }
\end{table*}

\begin{table*}
  \caption{Comparison of assessor models, Cohen's Kappa and F1 score}
  \label{tab:assessor_comparison}
  \centering
  \resizebox{0.9\textwidth}{!}{
  \begin{tabular}{lcccc}
    \hline
    Model Name & DeepSeek-VL & Qwen2-VL-72B-Instruct & Llava-OneVision & GPT-4o-mini \\
    \hline
    Cohen's Kappa & 0.00 & 0.36 & \textbf{0.43} & 0.32 \\
    F1 score & 55.9 & \textbf{78.0} & \textbf{78.0} & 76.3 \\
    \hline
  \end{tabular}
  }
\end{table*}

Our study evaluates multiple LVLMs without relying on domain-specific data. We assess the following models: Qwen2-VL-7B-Instruct~\cite{qwen2-vl}, Llama-3.2-11B-Vision-Instruct~\cite{llama-vision}, GPT-4o-mini, MedGemma-4B, MedGemma-27B~\cite{medgemma}, and Gemma-27B~\cite{gemmateam2025gemma3technicalreport}, though an arbitrary LVLM may be used in our benchmark. Including open-source models provides insight into the baseline capabilities of publicly available systems for diagnostic tasks. At the same time, adding MedGemma enables a direct comparison between general-purpose and medically specialized LVLMs, highlighting the effect of domain-specific adaptation.


We evaluate six prompting variants for GPT-4o-mini to systematically and equally study the contribution of visual and textual modalities to diagnostic accuracy and to analyze the effect of different prompting strategies. The first two dialogue-free options provide the lower-bound (\textbf{Image + General Complaint}) and upper-bounds (\textbf{Image + All Complaints}) for immediate diagnosis $\hat{D}_k$ from the image $P_k$, general complaint $BC_k$, and, in the latter case, additional complaints $AC_k$. Next, we examine various dialogue options: \textbf{Dialogue Only} diagnosis from the dialogue $DI_k$, without access to image $P_k$, \textbf{Dialogue + Image} with image $P_k$ included during the dialogue (also used for other model families), and \textbf{Dialogue + Image + Rationale} with rationale generation, in which the Doctor Agent explains each step of reasoning internally (hidden from the patient), promoting logical consistency~\cite{weietalcot}. Finally, we examine the possibility (\textbf{Dialogue + Image + Rationale + External Cues}) to combine LVLM with a ConvNet fine-tuned on the 3MDBench image training set (see details in Appendix~\ref{sec:baselines_cnn}), in which we add top-3 classes, predicted by the ConvNet from image $P_k$. Appendix~\ref{prompt-doctor-simple} provides prompt templates for each setup. 
\section{Results}

\begin{table*}[h!]
  \caption{Main results of our benchmark: diagnostic F1 scores of doctor agent}
  \label{tab:f-score}
  \resizebox{\textwidth}{!}{
  \begin{tabular}{lccc}
    \hline
    Model Name & Configuration & F1 Score & Number of utterances \\
    \hline
    EfficientNetV2-XL & Fine-tuned on the train part & 61.0 & - \\
    \hline
    GPT 4o-mini & No dialogue, image + general complaint  & 50.4 & - \\
    & No dialogue, image + all complaints & 66.8 & -\\
    & Dialogue, no image & 52.8 & 15.22 (±3.63) \\
    & Dialogue + image & 54.2 & 13.32 (±3.33) \\
    & Dialogue + image + rationale & 56.9 & 14.99 (±4.23) \\
    & Dialogue + image + rationale + external cues  & \textbf{70.3} & 14.48 (±3.97)\\
    \hline
    Llama-3.2-Vision & Dialogue + image & 41.5 & 14.49 (±4.02) \\
    \hline
    Qwen2-VL & Dialogue + image & 39.0 & 15.11 (±4.39) \\
    \hline
    MedGemma-4B & Dialogue + image & 37.9 & 17.48 (±4.84) \\
    \hline
    MedGemma-27B & Dialogue + image & 45.7 & 16.88 (±5.25) \\
    \hline
    Gemma3-27B & Dialogue + image & 51.1 & 14.81 (±3.81) \\
    \hline
  \end{tabular}
  }
\end{table*}

To enhance the reproducibility of 3MDBench while maintaining dialogue variability, we configured the Patient Agent with a maximum of 256 new tokens and a temperature of 0.6 and the Doctor Agent with a maximum of 512 completion tokens and a temperature of 0.6~\cite{gusev2025pingpongbenchmarkroleplayinglanguage}. To ensure stability in assessment, we set the Assessor Agent with a maximum of 512 new tokens and a temperature of \(1 \times 10^{-6}\). 
To ensure the statistical testing process, we employed the Wilcoxon signed-rank test with a significance level of $\alpha=0.01$ to assess the statistical significance of the difference in evaluated metrics. We applied false discovery rate control using the Benjamini–Hochberg procedure to account for multiple comparisons~\cite{benjamini1995controlling,hochberg2009multiple,pmlr-v202-savchenko23a}.

\subsection{Patient and Assessment Agents}

We evaluated the first two metrics from Section~\ref{subsec:patient} using GPT-4o-mini. Table \ref{tab:patient_models} presents the metrics on the validation set of 3MDBench. To make the final selection, we calculated the mean rank for each model across each metric and then averaged them. As a result, we chose Llama-3-8B as our patient model to ensure the benchmark remains open-access and independent of proprietary models. Moreover, by this selection, we implemented one of the proposed hypotheses for paraphrasing text to inhibit self-recognition, thereby mitigating the risk of employing the same model (GPT-4o-mini) for both Doctor Agent and symptom generation~\cite{panickssery2024llmevaluatorsrecognizefavor}.

To estimate the dialogue closeness, we calculated Cohen's Kappa for each criterion from Table~\ref{tab:doctor_criteria} and averaged the scores to determine overall agreement. For the diagnosis extraction task, we computed the F1 score of the diagnoses identified by the LVLM, using human-extracted diagnoses as the ground truth. Table \ref{tab:assessor_comparison} presents the evaluation results. 
Based on Cohen's Kappa and F1 score, we selected Llava-OneVision-Qwen2-72b-ov-chat-hf as the final Assessor Agent model. The agreement level of this model with the human annotation ($\kappa$=0.43) is similar to the inter-annotator agreement ($\kappa$=0.49), indicating that the model captures domain-relevant judgment rather than producing arbitrary outputs, which is consistent with recent findings on LLMs in clinical assessment~\cite{Kornblith2025Analyzing}.


\subsection{Diagnostic Results: Doctor Agent}

\begin{table*}[h!]
  \caption{Clinical competence of dialogue doctor systems. See details for criteria in Table~\ref{tab:doctor_criteria}}
  \label{tab:clin-competence}
  \begin{center}
  \resizebox{0.95\textwidth}{!}{
  \begin{tabular}{lccccccccc}
    \hline
    Model & 1.1 & 1.2 & 1.3 & 2.1 & 2.2 & 3.1 & 3.2 & 4.1 \\
    \hline
    GPT, dialogue, no image & \textbf{1.0} & \textbf{1.0} & 0.95 & \textbf{1.0} & \textbf{1.0} & 0.89 & 0.90 & 1.45 \\
    GPT, dialogue + image & 0.99 & \textbf{1.0} & 0.96 & \textbf{1.0} & \textbf{1.0} & 0.90 & 0.91 & 1.61 \\
    GPT, dialogue + image + rationale & 0.96 & 0.99 & 0.89 & 0.99 & 0.97 & 0.78 & 0.78 & 1.31 \\
    GPT, dialogue + image + rationale + external cues & 0.96 & 0.99 & 0.94 & 0.99 & 0.98 & 0.88 & 0.88 & 1.47 \\
    Llama-3.2-Vision & 0.99 & 0.99 & 0.96 & 0.99 & 0.99 & 0.75 & 0.74 & 1.45 \\
    Qwen2-VL & 0.90 & 0.93 & 0.78 & 0.92 & 0.90 & 0.61 & 0.61 & 1.16 \\
    MedGemma-4B & 0.97 & 0.98 & 0.94 & 0.99 & 0.98 & 0.79 & 0.80 & 1.42 \\
    MedGemma-27B & \textbf{1.0} & \textbf{1.0} & \textbf{1.0} & \textbf{1.0} & \textbf{1.0} & 0.90 & 0.88 & \textbf{1.67} \\
    Gemma3-27B & 0.99 & \textbf{1.0} & 0.99 & \textbf{1.0} & \textbf{1.0} & \textbf{0.97} & \textbf{0.98} & 1.57 \\
    \hline
  \end{tabular}
  }
  \end{center}
\end{table*}

A core assumption of our benchmark is that qualitative dialogue and medical imaging improve diagnostic performance (Table~\ref{tab:f-score}). Dialogue raises GPT-4o-mini’s F1 from 50.4 (image + basic complaint \( BC_k \)) to 54.2\% (p < 0.01), surpassing open-source models but still below the 66.8\% score of an unreal full-information setting where all patient details are known~\cite{li2024mediqquestionaskingllmsbenchmark}. This gap shows that LVLMs often miss key symptoms or end questioning prematurely, unlike human clinicians who adaptively probe. Stronger dialogue strategies are therefore needed to better approximate full-information performance.


Second, the results demonstrate that dialogue quality depends on access to image inputs. Here, with improving GPT-4o-mini F1-score from 52.8 to 54.2\% (p < 0.01), the average number of utterances per dialogue decreased from 15.22 (±3.6) without image access to 13.32 (±3.3) with image access (p < 0.01). Hence, the inclusion of visual information not only improves diagnostic accuracy but also leads to shorter, more efficient interactions.

We tested prompting strategies that avoid direct fine-tuning to demonstrate the effect of various strategies for the Doctor Agent. Building on prior work suggesting the benefits of chain-of-thought prompting~\cite{weietalcot}, rationale generation shows significant GPT-4o-mini F1 improvement over standard dialogue (56.9\% vs. 54.2\%, p < 0.01), indicating that explanations alone can enhance diagnostic reasoning in complex tasks. Moreover, enriching input with image-based cues, specifically the top-3 predictions from a fine-tuned EfficientNetV2-XL~\cite{pmlr-v139-tan21a} with details in Appendix~\ref{sec:baselines_cnn}, boosts the model's F1 score to 70.3\%, outperforming the full-information setting and EfficientNetV2-XL-only (p < 0.01). Thus, integrating a domain-specific vision model with a general-purpose LVLM may significantly improve the diagnostic ability.

\begin{figure}[h]
  \centering
  \includegraphics[width=\linewidth]{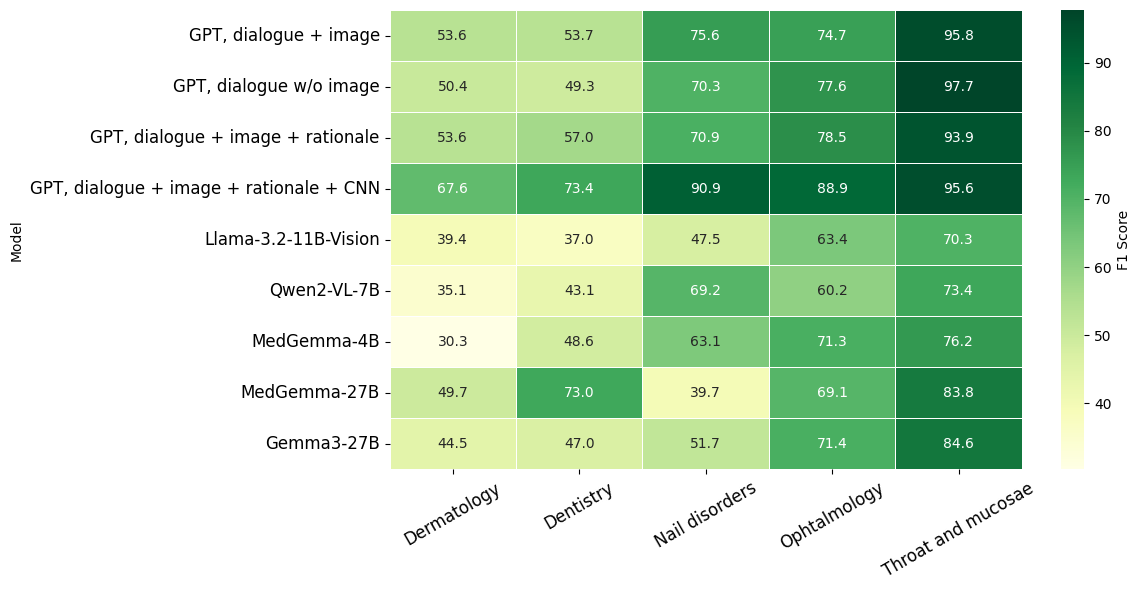}
  \caption{F1 scores by diagnosis categories}
  \label{fig:f1-categories}
\end{figure}

Across model families, 3MDBench shows a clear hierarchy of diagnostic ability. In the dialogue + image setting, GPT-4o-mini leads with 54.2\%, ahead of all open-source models. Among these, Gemma-27B performs best (51.1\%), surpassing Llama-3.2-Vision (41.5\%) and Qwen2-VL (39.0\%). Domain-specialized MedGemma models show gains with scale—MedGemma-4B at 37.9\% versus MedGemma-27B at 45.7\%—yet still lag behind Gemma-27B despite their medical focus. This gap likely reflects fine-tuning bias: its multimodal pretraining concentrates on clinical imaging—particularly chest X-rays and CT/MRI slices—alongside other specialist modalities (large-scale histopathology patches and retinal fundus images), with hundreds of thousands of radiology examples, while everyday photographs and lay symptom depictions are comparatively underrepresented~\cite{sellergren2025medgemmatechnicalreport}. Consequently, its inductive bias favors specialist modalities over common outpatient complaints, limiting effectiveness in telemedicine consultations and constraining diagnostic accuracy.


We also evaluated diagnostic accuracy across five disease categories, as shown in Figure~\ref{fig:f1-categories}. Performance varies considerably by category. Dermatology, with many overlapping conditions, yields the lowest average F1 (47.1\%), while throat/mucosae, with more apparent distinctions, scores highest (85.7\%). This result reflects model limitations in fine-grained classification and the dataset's uneven diagnostic coverage.

\subsection{Benchmarking Clinical Competence}

\begin{figure*}[ht]
  \centering
  \includegraphics[width=0.7\linewidth]{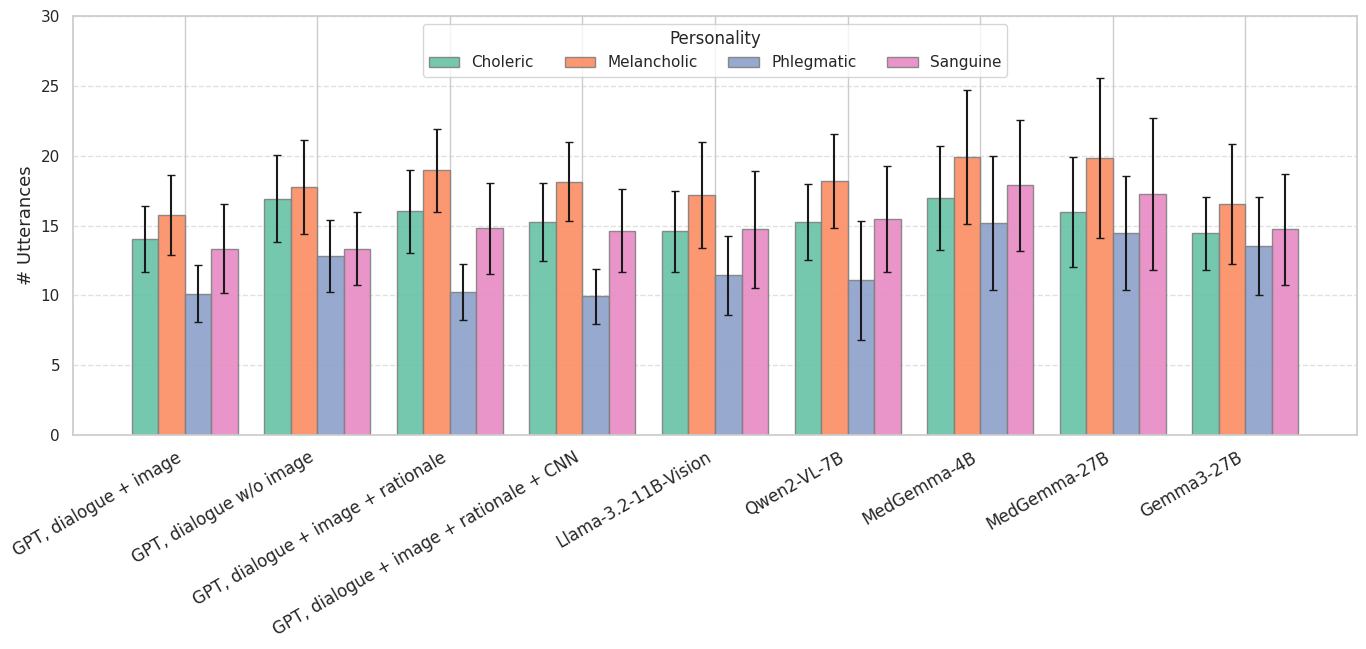}
  \caption{Number of utterances by personality types}
  \label{fig:utterances-pers}
\end{figure*}

Beyond diagnostic accuracy, we evaluated general clinical competence using Table~\ref{tab:doctor_criteria} criteria. As shown in Table~\ref{tab:clin-competence}, GPT- and MedGemma-based models achieve consistently high scores, with no criterion below 0.78, indicating strong adherence to professional and patient-oriented communication standards. \textit{Humanistic Care} (2.1–2.2) approaches saturation for stronger models, reflecting a ceiling effect on fundamental human-centered communication skills. \textit{Medical Interviewing Skills} (1.1–1.2) are likewise high, but reveal systematic differences by modality: without visual input, models compensate by probing more actively into patient history and symptoms, often extending questioning to reduce diagnostic uncertainty.

In contrast, the clinically demanding \textit{Diagnostic and Treatment Abilities} (3.1–3.2) remain strongly discriminative, separating GPT and MedGemma-based agents from open-source models such as Qwen2-VL. Further, GPT-4o-mini exhibits a trade-off: rationale-free variants score higher on communication skills, while rationale-based versions lag, suggesting inward reasoning can detract from applied clinical performance. Competence also varies by patient temperament (Figure~\ref{fig:clin-comp-pers}, Appendix~\ref{app:performance_analysis}); phlegmatic personas lower scores, indicating that Mini-CEX criteria still expose context-dependent weaknesses even in otherwise competent models.

\subsection{Patient Temperament}

\begin{figure}[h!]
  \centering
  \includegraphics[width=1.0\linewidth]{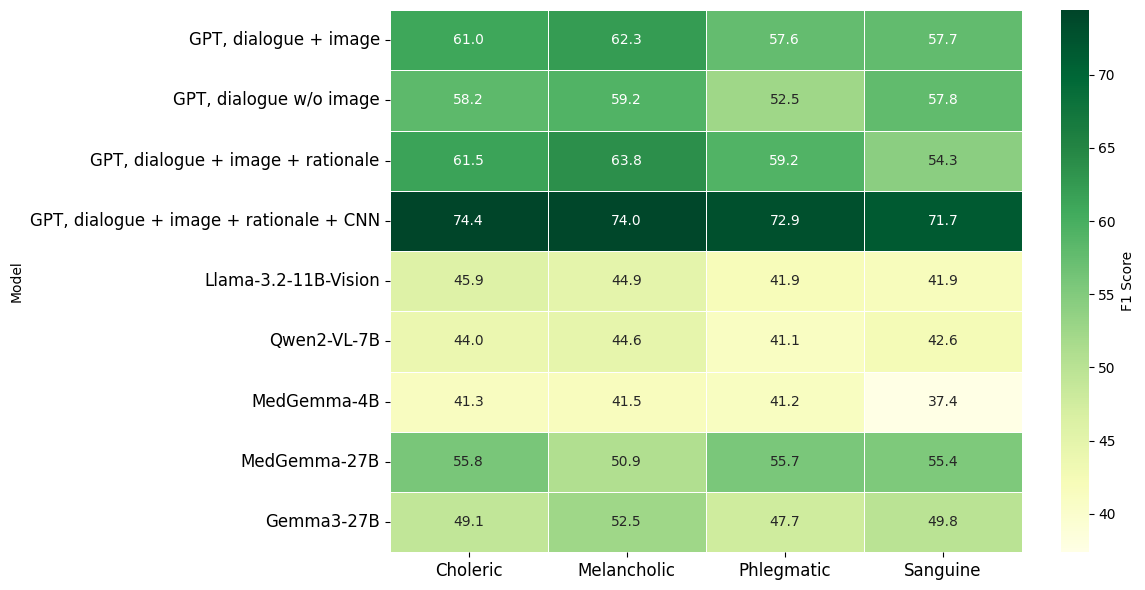}
  \caption{F1 scores by personality types}
  \label{fig:f1-score-pers}
\end{figure}

Our experiments with personality types (Figures~\ref{fig:f1-score-pers}, \ref{fig:utterances-pers}, see also detailed results in Figure~\ref{fig:clin-comp-pers}) show no statistically significant differences in F1 score between personalities. Indeed, LVLMs can maintain coherent, goal-directed dialogue even when faced with challenging behaviors, such as the sanguine patient's digressions or the melancholic patient's tendency to ask rather than answer questions.

Dialogues with phlegmatic patients yield slightly lower F1 and competence scores. Their short, passive responses limit the model's ability to gather rich clinical information, forcing the doctor to ask more follow-up questions. For example, as shown in Appendix~\ref{app:dial}, when asked about new product exposure, a phlegmatic patient may simply reply ``No''. By contrast, sanguine patients often provide unsolicited details, such as changes in daily routines or symptom triggers, that more effectively guide diagnosis.


Moreover, as shown in Figure~\ref{fig:utterances-pers}, dialogues with phlegmatic patients are, on average, four turns shorter due to their tendency not to ask clarifying questions. Appendix~\ref{app:dial} reveals that the doctor agent rarely compensates for this brevity by steering the conversation or probing deeper. While diagnostic accuracy is generally maintained, these interactions result in fewer recommendations or explanations.

Thus, although LVLMs are robust to diverse user behaviors, their performance may still degrade with minimally cooperative patients. This observation underscores the need to assess models' initiative and adaptability in less cooperative settings.
\section{Conclusion}

This paper introduces 3MDBench, an open-source benchmark for evaluating LVLMs in medical diagnostics. It simulates interactive telemedicine consultations, incorporating diverse diagnoses and patient behaviors to assess diagnostic accuracy and clinical competence. 

Within this framework, we demonstrate that the ability to engage in dialogue with the patient and the visual modality significantly enhances diagnostic accuracy. General-purpose LVLMs display strong clinical competence, effectively leveraging images and conducting information-seeking dialogues to provide accurate diagnoses.

We highlight a key limitation of LVLMs in medical diagnostics: while strong in human-centered communication, they lack domain-specific visual expertise. We demonstrate how to improve the quality of Doctor Agent using our benchmark by incorporating predictions from a convolutional network trained on the diagnosis prediction task, significantly (up to 20\%) enhancing LVLM performance. Thus, combining general-purpose LVLMs with lightweight, task-specific vision models offers a scalable opportunity to higher performance without extensive supervised fine-tuning.

Our findings suggest that while dialogue contributes to more accurate diagnosis, its effectiveness is limited. External expert cues and better prompting can bridge the gap, while broader and more balanced diagnostic coverage remains a key goal for future benchmarks.

\section*{Limitations}

\paragraph{Using LLMs for symptom generation and assessment}

We generated additional patient symptoms using GPT-4o-mini, conditioned on the image and diagnosis. Although this approach leverages embedded medical knowledge and was partially validated by physicians on the validation split, it may still introduce factual inaccuracies or distributional biases. Similarly, LLM-based evaluation under the LLM-as-a-judge paradigm~\cite{Zheng2023JudgingLW} depends on the assessor's domain competence and may propagate systematic imperfections. We conducted manual checks for plausibility and coherence, but we cannot guarantee absolute correctness.

\paragraph{Dataset coverage and diagnostic constraints}

The benchmark uses publicly available images, which introduces a potential risk of data leakage from pretraining. The current set of 34 diagnoses, although selected to reflect consultation distributions and curated with attention to data availability and assessment reliability, still provides limited diagnostic coverage. In addition, the Doctor Agent must choose a single diagnosis from this predefined set, which ensures comparability but reduces clinical realism. Future iterations should expand the disease set, incorporate free-text outputs with mapping to ICD-10/UMLS~\cite{Bodenreider2004-pv}, or hierarchy-aware scoring to support more open-ended evaluation.

\paragraph{Patient simulation}

While the four temperament categories provide a valuable foundation for simulating patient diversity, future work could explore more nuanced or data-driven patient behavior models to more accurately reflect the complexity observed in real-world clinical interactions.
\section*{Ethics Statement}

\paragraph{Human Involvement}

This work involved several instances of human annotation. First, one of the co-authors with a medical background reviewed the collected images over one week to verify the correctness of the associated diagnoses. Second, we obtained human annotations to evaluate dialogues for selecting the Assessor Agent. Five employees completed the annotation process, each dedicated approximately six hours to the task during their regular working hours, without additional compensation. All annotators were informed of the research purpose behind the annotation tasks.

\paragraph{Inference Costs}

Running the complete evaluation experiment on a single A100 GPU took approximately 48 hours to select the Patient Agent model, 4 hours to select candidate Assessor Agent models, and 210 hours to evaluate the Doctor Agents.

\paragraph{Use of AI Assistants}

We used Grammarly to improve and proofread the text of this paper, including grammar, spelling, style corrections, and sentence rephrasing. As a result, some parts of the manuscript may be classified as AI-generated, AI-edited, or a mix of human and AI contributions.

\section*{Acknowledgments}
The work of A. Savchenko was supported by a grant, provided by the Ministry of Economic Development of the Russian Federation in accordance with the subsidy agreement (agreement identifier 000000C313925P4G0002) and the agreement with the Ivannikov Institute for System Programming of the Russian Academy of Sciences dated June 20, 2025 No. 139-15-2025-011.

\bibliography{acl_latex}

\appendix
\definecolor{promptbg}{RGB}{240, 240, 240} 
\definecolor{diagnosisblue}{RGB}{0, 0, 255} 
\definecolor{titlebg}{RGB}{200, 200, 200} 

\clearpage

\section{3MDBench Statistics}
\label{app:3mdbench_table} 

\begin{figure*}[h]
  \centering
  \includegraphics[width=\linewidth]{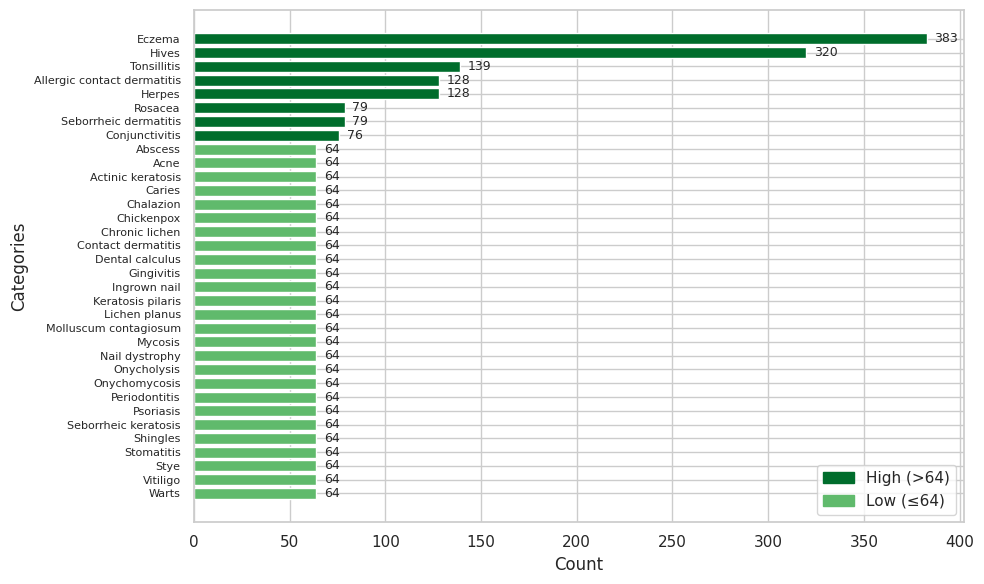}
  \caption{The distribution of classes in 3MDBench. The dataset consists of 34 medical conditions, with the most frequent class containing 383 samples, while 21 classes have exactly 64 samples (highlighted in light green).}
  \label{fig:data_distr}
\end{figure*}

\begin{figure*}[h]
  \includegraphics[width=1.\linewidth]{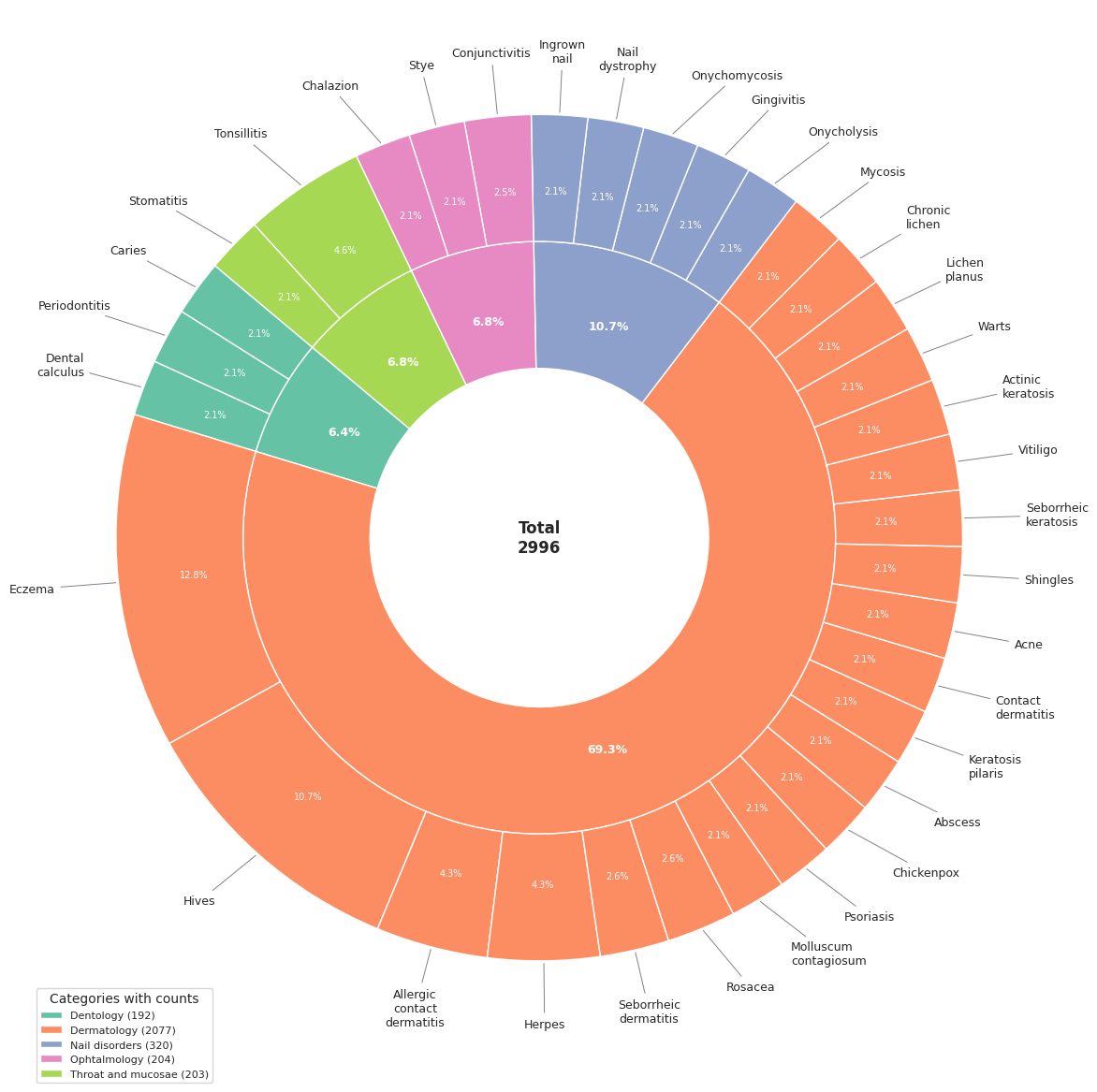}
  \caption{Distribution of selected diagnoses based on real-world telemedicine consultations.}
  \label{fig:real_distr}
\end{figure*}

Figure~\ref{fig:data_distr} shows the class distribution in 3MDBench, obtained by merging data from the utilized datasets and enriching them with data from the Bing Image Search API. We adjusted the resulting distribution to approximate real-world diagnosis frequencies observed in telemedicine consultation. 
Figure~\ref{fig:real_distr} presents the distribution of medical diagnoses derived from real-world telemedicine consultations and grouped by medical category.

\twocolumn

\section{Human Agreement}
\label{sec:agreement}

\begin{figure}[h]
  \centering
  \includegraphics[height=\linewidth]{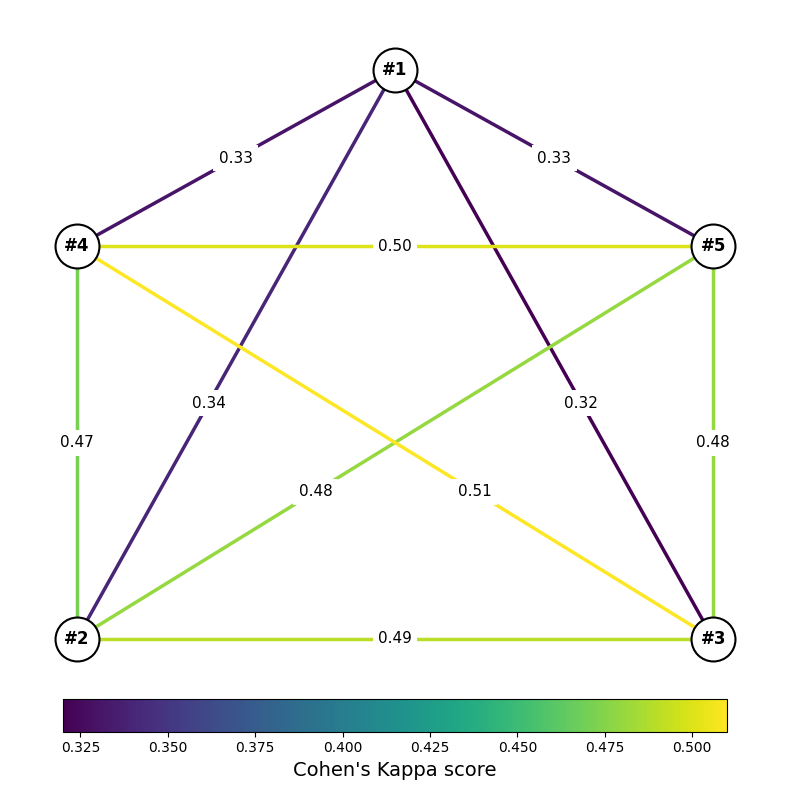}
  \caption{Inter-rater agreement graph showing pairwise Cohen’s Kappa scores between participants \#1–\#5.}
  \label{fig:kappas}
\end{figure}

We assessed inter-rater agreement by analyzing consistency across all participants. Annotation was conducted using Google Forms so annotators could not see each other's answers. Figure~\ref{fig:kappas} presents a graph with nodes representing anonymized participants (\#1 to \#5) and edge weights corresponding to pairwise Cohen's Kappa scores. Analysis of these values indicated that participant \#1 exhibited consistently lower agreement with the other annotators (average pairwise Kappa = 0.33), thereby reducing the overall mean Kappa to 0.42.

Further investigation revealed that this participant exhibited abnormally fast task completion times and random-like response patterns, suggesting noncompliance with instructions. After we excluded this data, the overall average Kappa across assessors increased to 0.49, indicating improved inter-rater reliability.

\section{Human Evaluation of Generated Patient Cases}
\label{sec:case_evaluation}

To ensure the quality and realism of the generated complaints, we validated them using human evaluators with relevant domain expertise. The evaluation tasks were distributed among physicians and advanced medical students. Each task was presented in a separate Google Form. Evaluators had no time constraints and were permitted to consult external sources to resolve doubts.
We evaluated the complaints in two distinct settings:
\begin{enumerate}
    \item \textbf{Basic Complaint Relevance:} Physicians assessed whether the primary complaint generated for each of the 34 diagnoses reflected the most typical and relevant symptom.
    \item \textbf{Contextual Complaint Relevance:} For one case per diagnosis (34 total), physicians assessed whether each additional complaint was relevant to the diagnosis and the image, providing contextually appropriate and visually grounded information supporting clinical decision-making.
\end{enumerate}
Each complaint was assigned a binary label (1 for relevant, 0 for not relevant). In the first task, at least one physician marked the basic complaint as relevant for $88\%$ of diagnoses. In the second task, at least one physician found the additional complaint relevant in $92\%$ of cases. We consider these inter-rater agreement rates acceptable for this task and consistent with prior literature on LLM response relevance in clinical settings~\cite{Singhal2023TowardsEM}.

\section{Human Expert Accuracy on Diagnostic Task}
\label{sec:human_accuracy}

To compare model and human diagnostic performance, we selected one case per diagnosis ($N=34$) and tasked four board-certified physicians with identifying the correct diagnosis under two experimental conditions:

\begin{enumerate}
    \item \textbf{Minimal information:} Presented only with the image and general complaint. This condition was evaluated by Physicians 1 and 2.
    \item \textbf{Full information:} Presented with the image, the general complaint, and the full list of generated symptoms. This condition was evaluated by Physicians 3 and 4.
\end{enumerate}

Each condition was assigned a different set of annotators to prevent cross-contamination and learning bias. The model's performance (GPT-4o-mini) was evaluated on the same set of cases under identical conditions for a direct comparison. The results are summarized in Table~\ref{tab:model_vs_human}.

\begin{table*}[h]
\centering
\caption{Performance comparison between the model (GPT-4o-mini) and human labelers across different settings. P-values are calculated against the model's F1 score.}
\label{tab:model_vs_human}
\begin{tabular}{l  l  c  c}
\hline
\textbf{Setting} & \textbf{Model / Labeler} & \textbf{F1 Score} & \textbf{p-value} \\
\hline
General complaint + image & GPT-4o-mini & 54.55 & -- \\
                         & Labeler \#1 & 39.39 & 0.22 \\
                         & Labeler \#2 & 27.27 & 0.02 \\
\hline
Full complaints list + image & GPT-4o-mini & 54.55 & -- \\
                            & Labeler \#3 & 57.58 & 0.81 \\
                            & Labeler \#4 & 54.55 & 1.00 \\
\hline
\end{tabular}
\end{table*}

Although this represents a small-scale pilot study, the results suggest that our LVLM can achieve diagnostic performance comparable to human physicians when provided with complete contextual information. This finding is consistent with a growing body of literature demonstrating that AI models can rival or even surpass medical experts in specific diagnostic tasks~\cite{Zeltzer2023Diagnostic,Tu2025Towards}.

\section{Baseline Computer Vision Convolution Model Selection}
\label{sec:baselines_cnn}  


To enhance LVLMs with disease-related information from images, we trained and compared different CV model of varying sizes to assess their performance in disease classification within the given setting. We selected the EfficientNetV2 model family, pre-trained on ImageNet-1k~\cite{russakovsky2015imagenetlargescalevisual} due to its efficient training process and competitive performance compared to other CNN and ViT models~\cite{pmlr-v139-tan21a}, and applied full fine-tuning on the train part of the 3MDBench. As shown in Table~\ref{tab:baseline_cnn_results}, the EfficientNetV2-XL model achieved the highest classification performance among the tested models. Therefore, we selected this model for further integration with the Doctor Agent.
\begin{table}
\centering
\caption{Performance comparison of baseline EfficientNetV2 models}\label{tab:baseline_cnn_results}
\begin{tabular}{ccccl}
\hline
\textbf{Model name} & \textbf{Macro F1 score, \%} \\ 
\hline
\textbf{EfficientNetV2 S} & 60 \\

\textbf{EfficientNetV2 M} & 52 \\

\textbf{EfficientNetV2 L} & 57 \\

\textbf{EfficientNetV2 XL} & 64 \\
\hline
\end{tabular}
\end{table}

\onecolumn

\section{Detailed 3MDBench Performance Statistics}
\label{app:performance_analysis}



\begin{figure*}[h!]
  \centering
  \includegraphics[width=\linewidth]{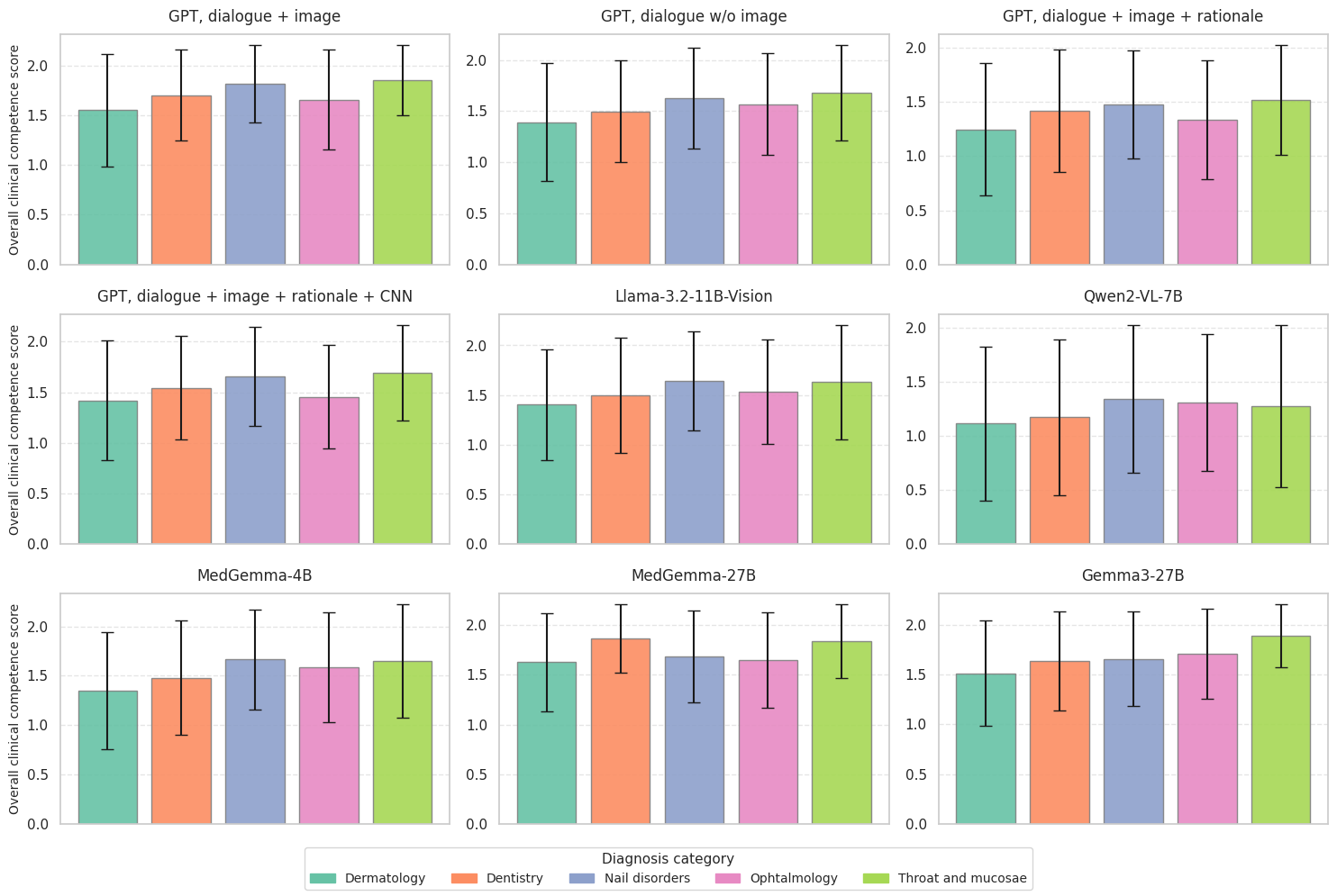}
  \caption{Overall clinical competence scores by diagnosis categories}
  \label{fig:clinical-competence-categories}
\end{figure*}

\begin{figure*}[h!]
  \centering
  \includegraphics[width=\linewidth]{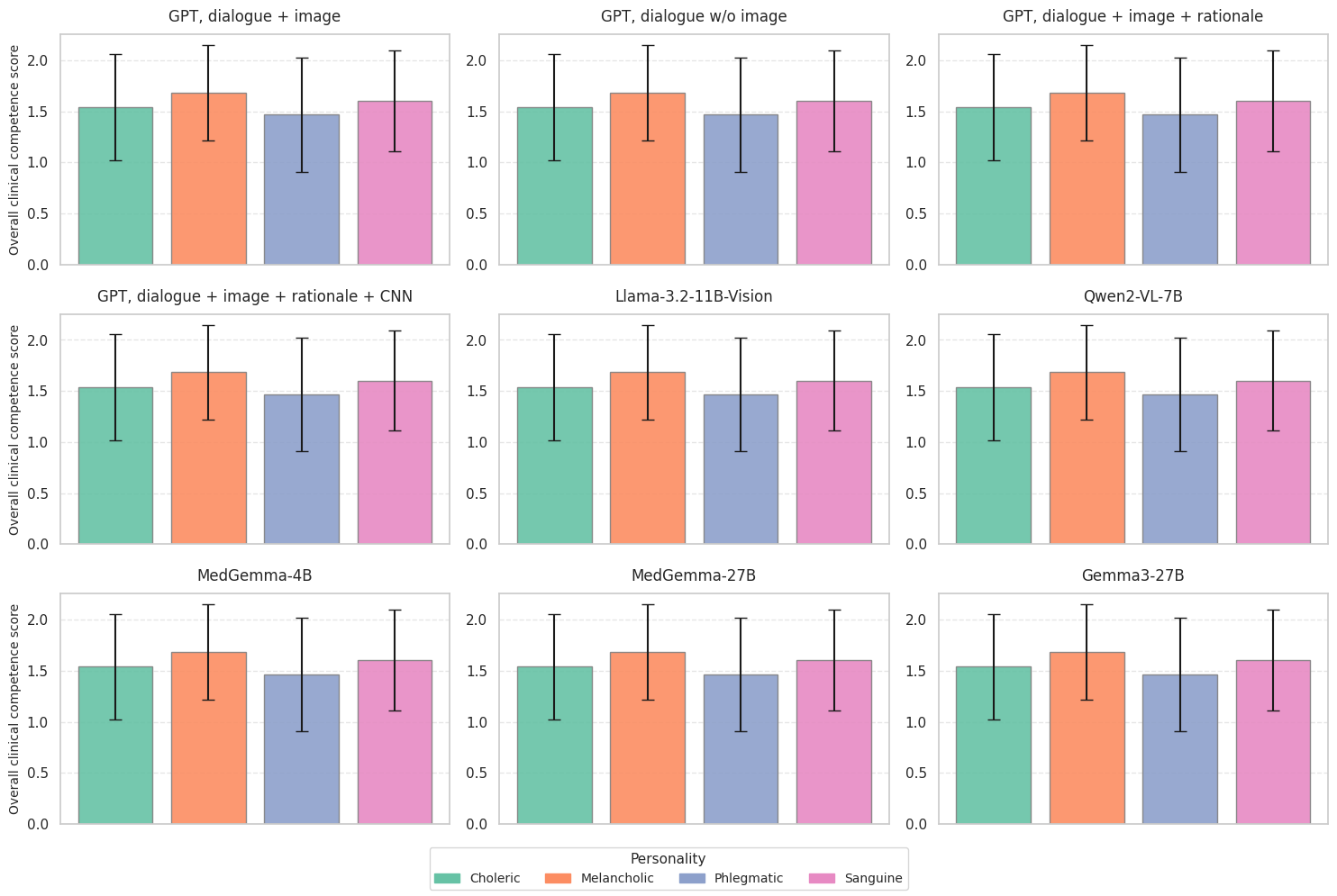}
  \caption{Overall clinical competence scores by personality types}
  \label{fig:clinical-competence-pers}
\end{figure*}

\begin{figure*}[h!]
  \centering
  \includegraphics[width=\linewidth]{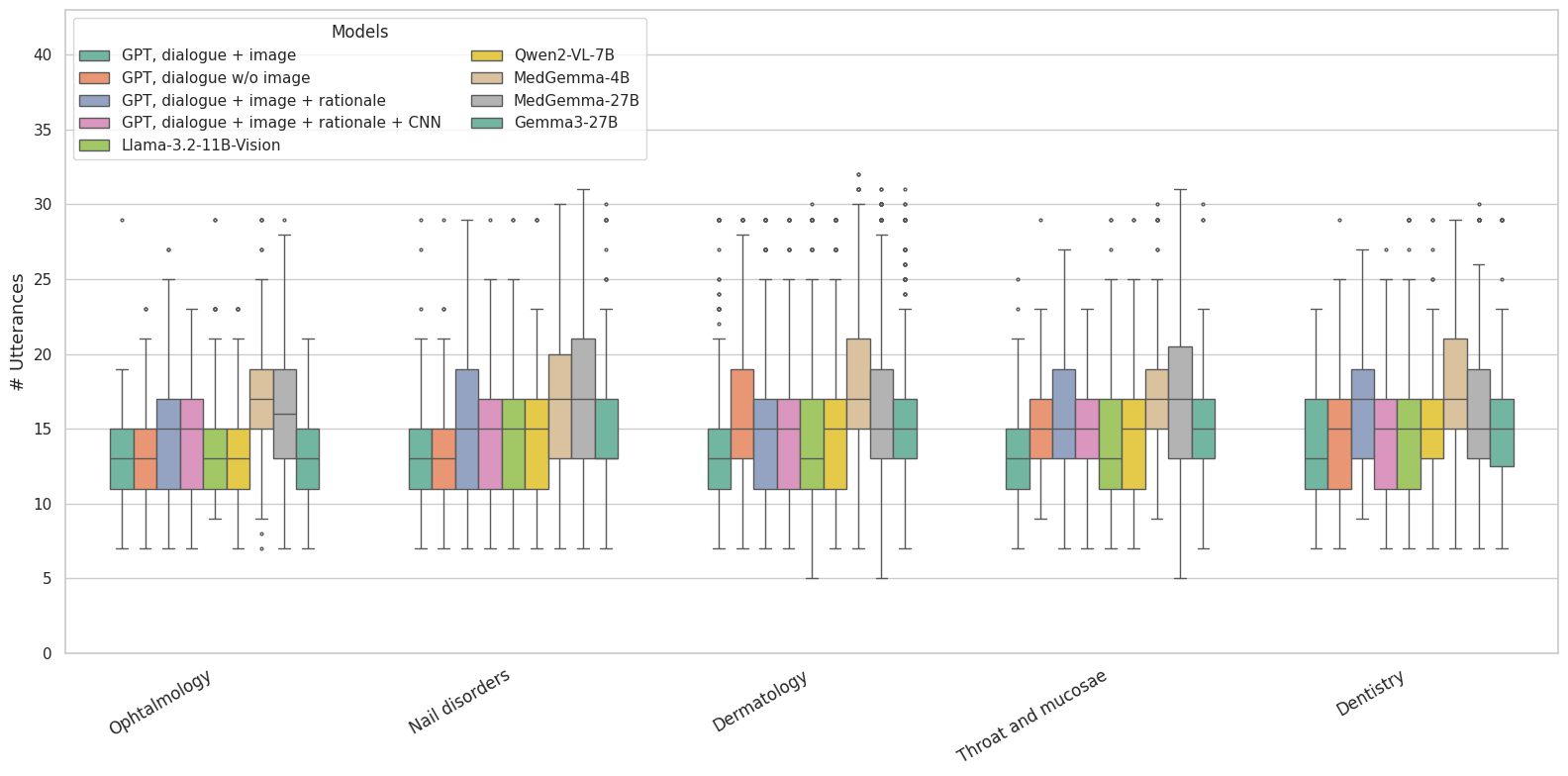}
  \caption{Number of utterances by diagnosis categories}
  \label{fig:utterances-categ}
\end{figure*}


\begin{figure*}[h!]
  \centering
  \includegraphics[width=\linewidth]{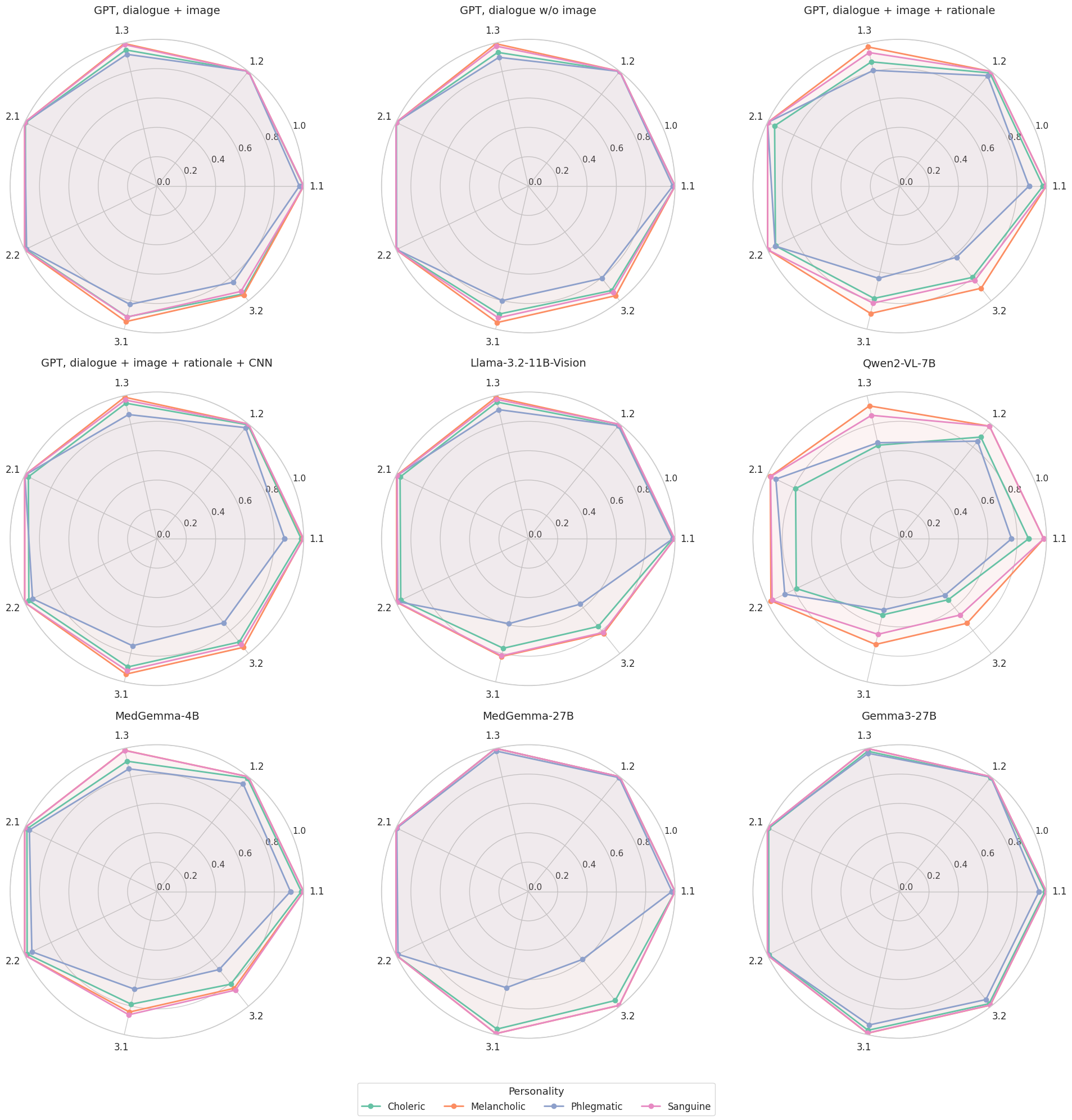}
  \caption{Clinical competence scores by personality types}
  \label{fig:clin-comp-pers}
\end{figure*}

\begin{figure*}[h!]
  \centering
  \includegraphics[width=\linewidth]{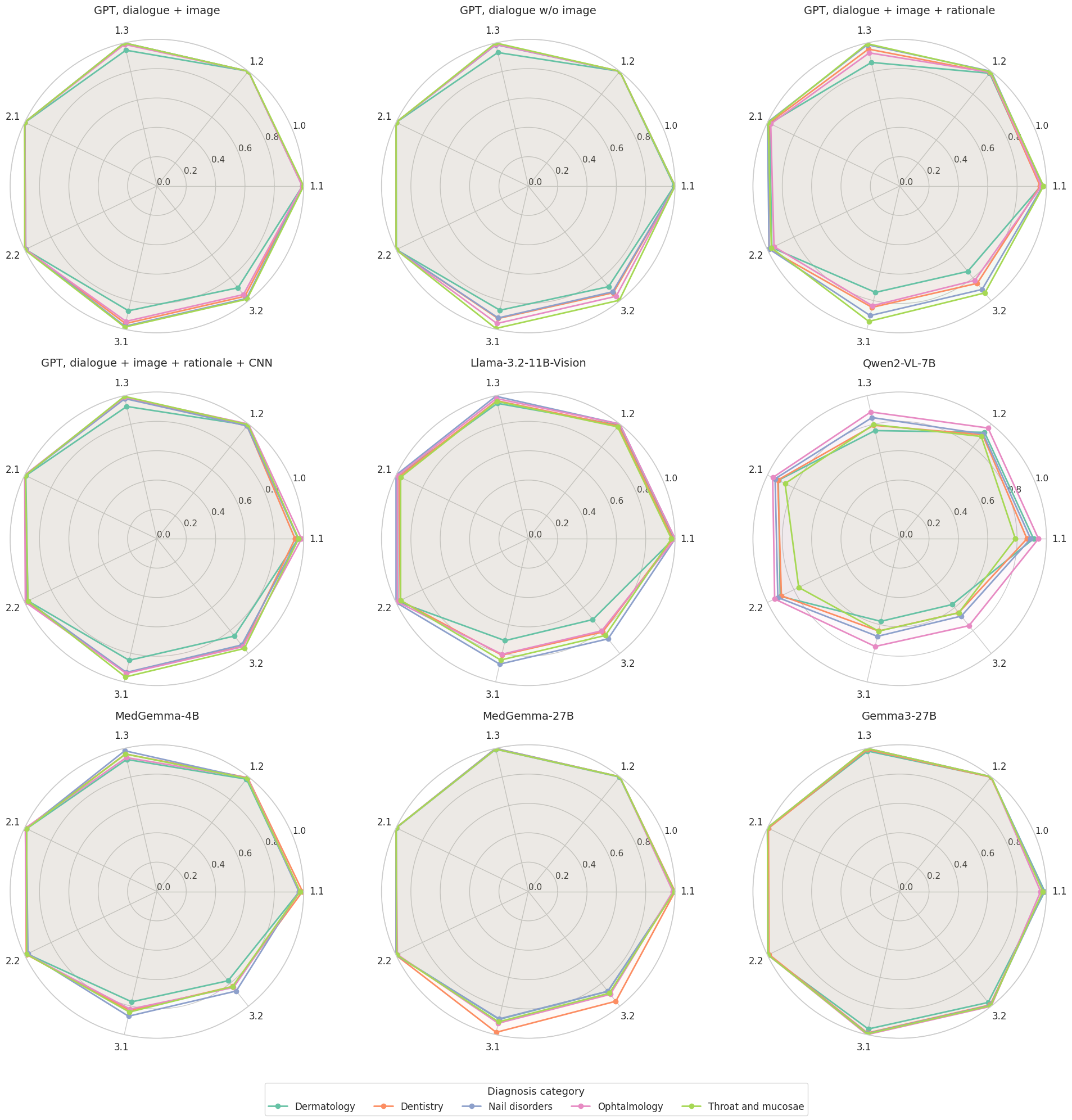}
  \caption{Clinical competence scores by diagnosis categories}
  \label{fig:clin-comp-categ}
\end{figure*}




\clearpage

\section{Diagnoses Prediction}
\label{app:diagnoses_prediction}

\begin{figure*}[h!]
  \centering
  \includegraphics[width=\linewidth]{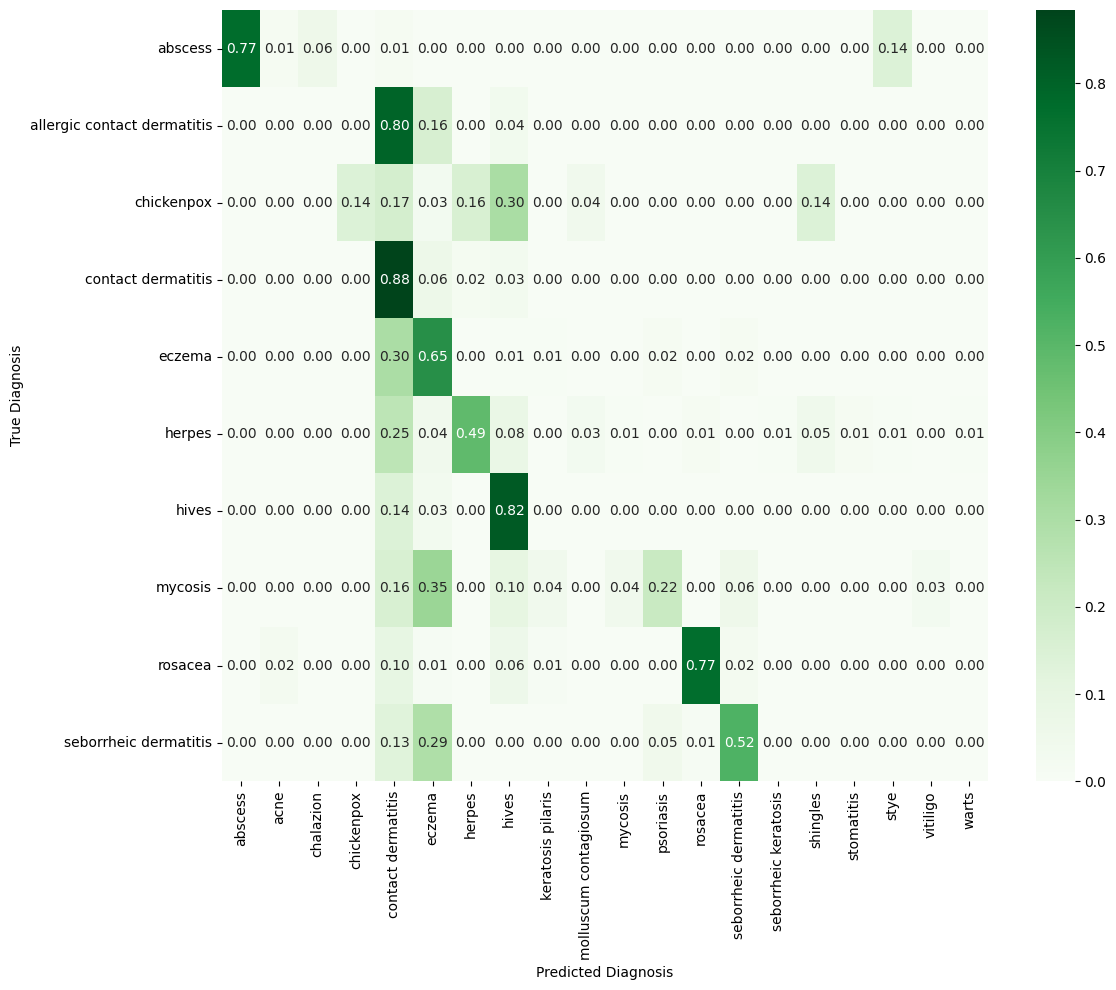}
  \caption{Confusion matrix for the predictions on the samples with the 10 most frequent diagnoses from the dermatology category for GPT-4o-mini with dialogue and rationale generation.}
  \label{fig:dermatology_diagnoses_cm}
\end{figure*}





\begin{figure}[h!]
  \centering
  \begin{subfigure}[b]{0.48\linewidth} 
    \includegraphics[width=\linewidth]{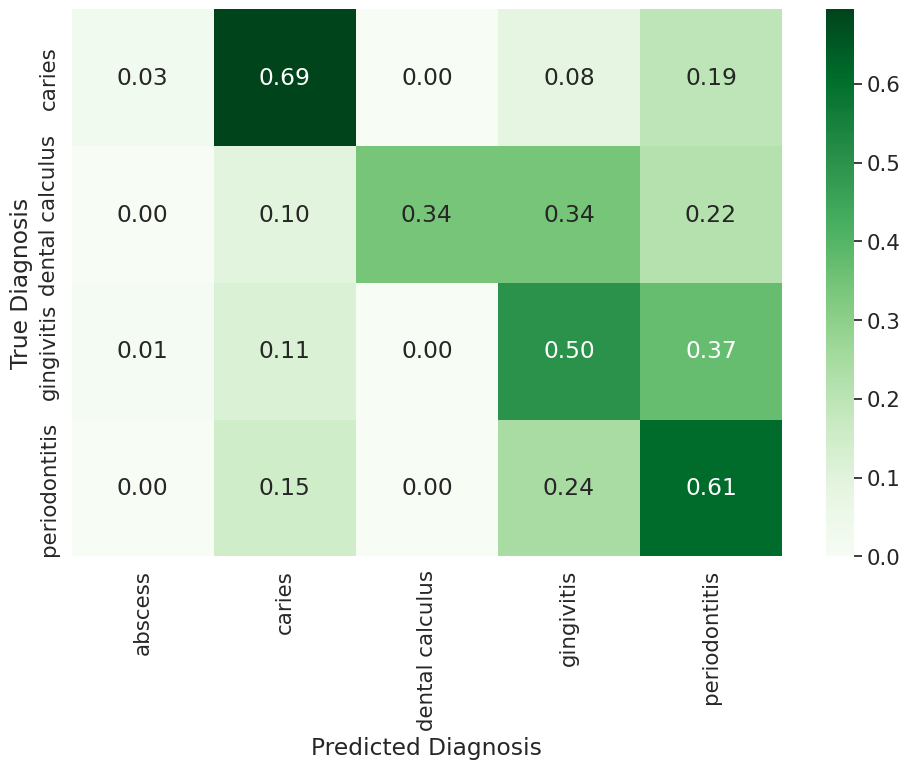}
    \caption{Dentistry category}
    \label{fig:dentistry_diagnoses_cm}
  \end{subfigure}
  \hfill 
  \begin{subfigure}[b]{0.48\linewidth}
    \includegraphics[width=\linewidth]{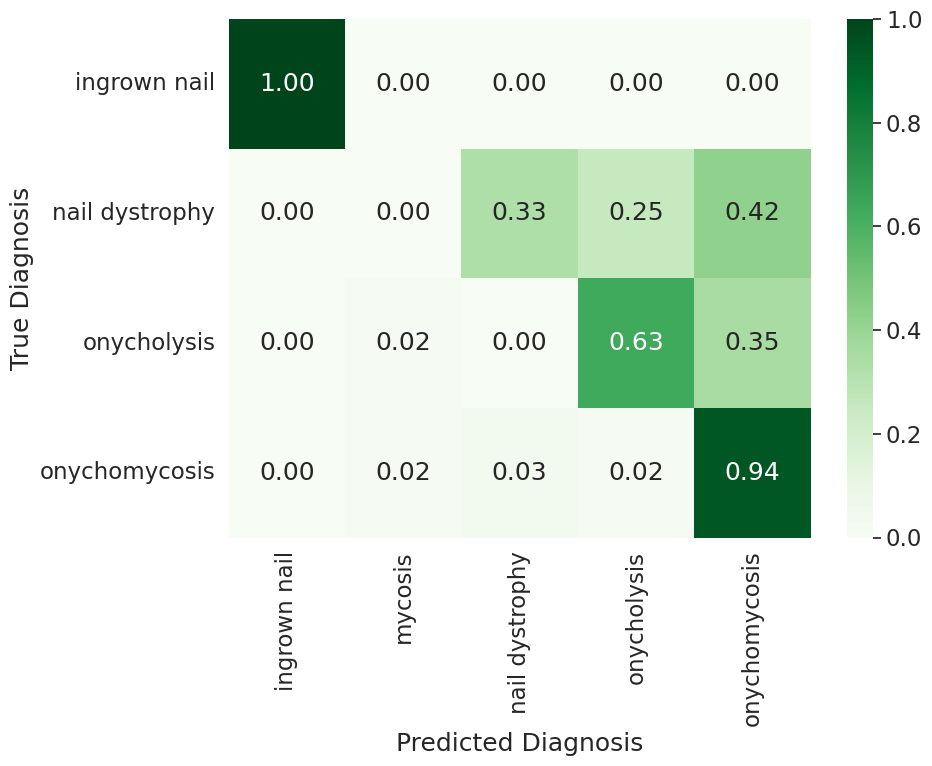}
    \caption{Nail diseases category}
    \label{fig:nails_diagnoses_cm}
  \end{subfigure}
  
  \vspace{0.5cm}
  
  \begin{subfigure}[b]{0.48\linewidth}
    \includegraphics[width=\linewidth]{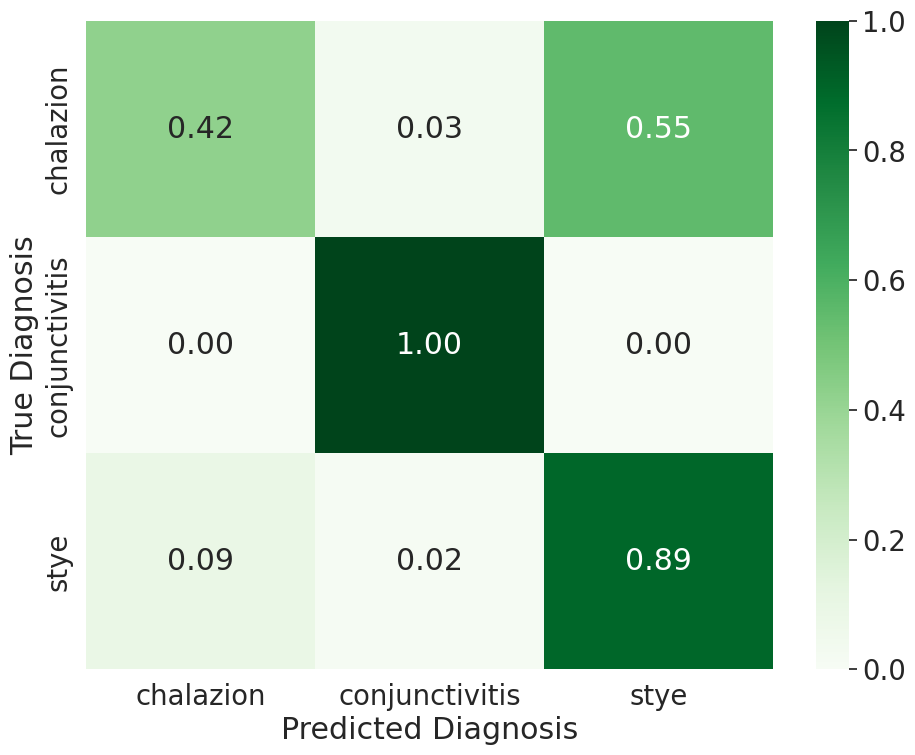}
    \caption{Ophthalmology category}
    \label{fig:ophtalmology_diagnoses_cm}
  \end{subfigure}
  \hfill
  \begin{subfigure}[b]{0.48\linewidth}
    \includegraphics[width=\linewidth]{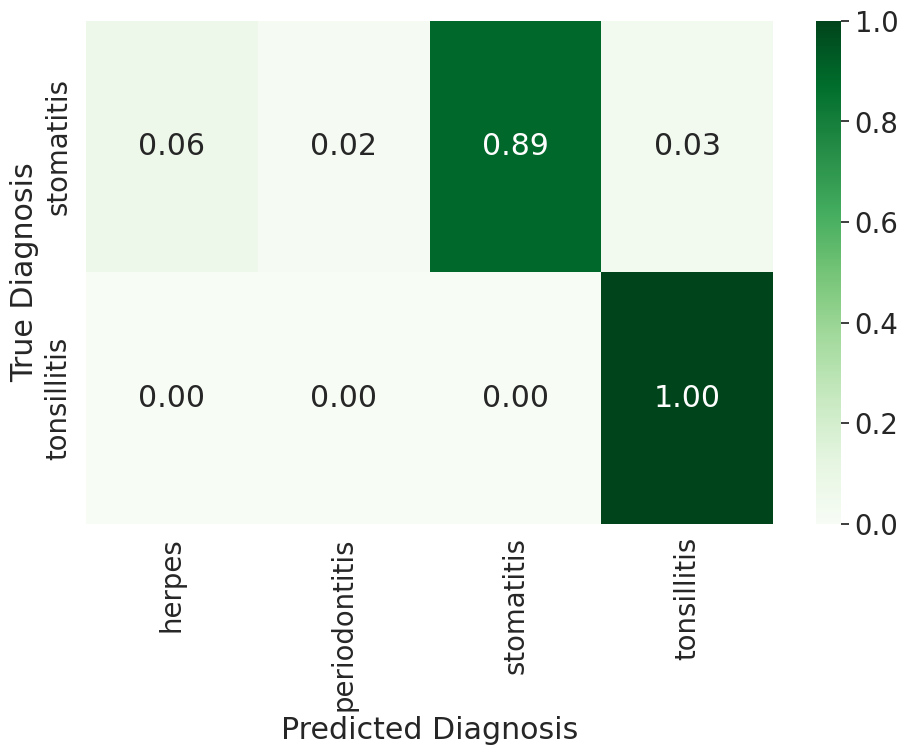}
    \caption{Throat and mucosae category}
    \label{fig:throat_diagnoses_cm}
  \end{subfigure}
  
  \caption{Confusion matrices for predictions by GPT-4o-mini with dialogue and rationale generation across different medical categories.}
  \label{fig:diagnoses_cm_grid}
\end{figure}

\onecolumn
\section{Prompts}

\subsection{Prompts for 3MDBench textual data generation}\label{app:text_gen}
\begin{center}
\begin{tcolorbox}[
    colback=promptbg, colframe=black, sharp corners=southwest, boxrule=0.5pt, width=\columnwidth,
    title={\textbf{Basic complaints generation prompt}}, coltitle=black, colbacktitle=titlebg, label={text-gen-base}
]
You are given a diagnosis. Assume the perspective of a human patient describing their personal experience in everyday language. \\
Please generate a single concise general symptom description that is most likely to occur for the given diagnosis. The description should be in the second person and contain at most 2 symptoms. \\

Example: \\
Diagnosis: eczema \\
Symptoms description: You have dry itchy patches on your skin. \\

Do not mention the diagnosis directly. Answer only with the description.

Diagnosis: \textcolor{diagnosisblue}{diagnosis name}
\end{tcolorbox}
\end{center}

\begin{center}
\begin{tcolorbox}[
    colback=promptbg, colframe=black, sharp corners=southwest, boxrule=0.5pt, width=\columnwidth,
    title={\textbf{Additional complaints generation prompt}}, coltitle=black, colbacktitle=titlebg, label={text-gen-add}
]
You are provided a diagnosis, corresponding photograph, and a general complaint. Assume the perspective of a human patient who recently discovered the symptoms and describes their personal experience to a doctor in everyday language.

Generate a complete yet concise medical case description. It must come in the form of unnumbered list of independent, atomic specific facts, each containing a single piece of information related to a patient experience in the context of the content of the photograph.
The list of complaints should add information to the given photograph.
The complaints in the list must not duplicate the general complaint; they should expand it. Do not repeat symptoms. Do not include emotional connotations.\\

The medical case must contain information about:

\hspace{1em} - The specific symptoms patient experiences (additionally to the ones already in the general complaint)

\hspace{1em} - The exact location of the symptoms and the area affected, if this can be inferred from the photograph

\hspace{1em} - How long the patient experiences the symptoms (if this is important for the diagnosis. this should be inferred from the photograph)

\hspace{1em} - How intense are the symptoms

\hspace{1em} - Were there any events that have led to the condition (only if this information is important for the diagnosis, otherwise, skip this)

\hspace{1em} - Does the patient have any known allergic reactions or chronic illnesses. If they usually do not cause the diagnosis, write that there are none.\\

Avoid technical or medical jargon. Each symptom must represent the smallest possible unit of information while remaining self-sufficient and focused on how the person might naturally describe their issues.

Do not repeat the symptoms and complaints. Do not mention anything that is already included in the general complaint. Do not add information about other accompanying illnesses.
Do not directly mention the diagnosis or use sub-lists. Write them in the second person.\\

Make sure that the complaints provided make it possible to provide the differential diagnosis. They should allow to distinguish the diagnosis from: \textcolor{diagnosisblue}{group name}

Diagnosis: \textcolor{diagnosisblue}{diagnosis name}

General complaint: \textcolor{diagnosisblue}{general complaint}
\end{tcolorbox}
\end{center}

\clearpage

\subsection{Prompts for Patient Agent selection}
\label{app:pat_select}






\begin{center}
\begin{tcolorbox}[
    colback=promptbg, colframe=black, sharp corners=southwest, boxrule=0.5pt, width=1\columnwidth,
    title={\textbf{Patient's Agent system prompt for Agent selection}}, coltitle=black, colbacktitle=titlebg, label={select-prompt-patient}
]

You are a patient at an online consultation with the general practitioner.

Your main complaint: \textcolor{diagnosisblue}{general complaint}

Additional symptoms: \textcolor{diagnosisblue}{symptoms}\\

Your goal in this conversation is to understand the cause of the symptoms, the diagnosis, and the treatment.
You cannot self-diagnose, you only tell the doctor about your symptoms.

RULES:

\hspace{1em} 1. When the doctor asks about symptoms, respond with ONLY ONE symptom that is MOST RELEVANT to their question.

\hspace{1em} 2. DO NOT mention more than one symptom in any answer.

\hspace{1em} 3. If you mention multiple symptoms, the response will be INVALID.\\

If the information the doctor inquires is not stated in your instructions, answer ONLY with ``I cannot answer this question, please do not ask this question again.''
Do not provide any analysis, inference, or implications.
Start the conversation with the main complaint ONLY.
If the doctor tells you the diagnosis and gives recommendations, respond with BREAK.
\end{tcolorbox}
\end{center}


\begin{center}
\begin{tcolorbox}[
    colback=promptbg, colframe=black, sharp corners=southwest, boxrule=0.5pt, width=1\columnwidth,
    title={\textbf{Doctor's prompt for Patient Agent selection confidence level}}, coltitle=black, colbacktitle=titlebg, label={select-prompt-doctor}
]

You are a highly experienced general practitioner with extensive medical expertise. 
Your task is to diagnose the patient's condition. \\

After every patient statement, you must evaluate your confidence level in diagnosing the condition using the following scale:

\hspace{1em} Unconfident: The evidence is minimal or inconclusive, making it impossible to reliably identify a diagnosis at this stage. Further investigation or data is necessary to differentiate between potential diagnoses.

\hspace{1em} Neutral: There is some evidence pointing toward a potential diagnosis, but it is insufficient to confidently identify the correct one. Additional information is needed to clarify.

\hspace{1em} Confident: The diagnosis is supported by available evidence, with a strong likelihood of being correct. While more information could solidify the decision, it is not strictly necessary for a high degree of certainty. \\

Analyse previous conversation carefully and make a decision.       
Answer in the following format:

\hspace{1em} CONFIDENCE LEVEL: [chosen rating from the list].

Write ONLY a confidence level.

\end{tcolorbox}
\end{center}

\begin{center}
\begin{tcolorbox}[
    colback=promptbg, colframe=black, sharp corners=southwest, boxrule=0.5pt, width=1\columnwidth,
    title={\textbf{Doctor Agent's prompt for Patient Agent selection; enquire information}}, coltitle=black, colbacktitle=titlebg, label={select-prompt-doctor-ask}
]
You are a highly experienced general practitioner with extensive medical expertise. Determine that you do not yet have enough information, and consider which missing detail is the most important to ask the patient to provide the most helpful information toward a correct medical decision.

Reflect on the common questions asked in similar cases and formulate ONE question the best and most intuitive doctor would ask in this situation. Continue dialogue with this question.

Analyze the whole dialogue: if the patient says they cannot answer a question during the conversation, remember these questions and do not ask further questions on those topics. Instead, select different suitable topics for consultation.
\end{tcolorbox}
\end{center}

\begin{center}
\begin{tcolorbox}[
    colback=promptbg, colframe=black, sharp corners=southwest, boxrule=0.5pt, width=1\columnwidth,
    title={\textbf{Doctor Agent's  prompt for Patient Agent selection; diagnose}}, coltitle=black, colbacktitle=titlebg, label={select-prompt-doctor-diagnose}
]
You are a highly experienced general practitioner with extensive medical expertise. You are confident that you have gathered sufficient information to make an informed decision, provide a diagnosis that aligns with the symptoms and history provided by the patient. 
Include a detailed explanation of the potential causes and offer tailored recommendations for treatment or further steps. \\

You must choose ONLY ONE diagnosis STRICTLY from the following list:

\texttt{['eczema', 'hives', 'herpes', 'allergic contact dermatitis', 'contact dermatitis', 'seborrheic dermatitis', 'rosacea', 'conjunctivitis', 'tonsillitis', 'chronic lichen', 'keratosis pilaris', 'molluscum contagiosum', 'lichen planus', 'periodontitis', 'caries', 'psoriasis', 'ingrown nail', 'stye', 'acne', 'onychomycosis', 'seborrheic keratosis', 'chalazion', 'vitiligo', 'shingles', 'mycosis', 'actinic keratosis', 'onycholysis', 'chickenpox', 'dental calculus', 'warts', 'stomatitis', 'abscess', 'gingivitis', 'nail dystrophy']}

Do not write about confidence level at the end.
\end{tcolorbox}
\end{center}


\begin{center}
\begin{tcolorbox}[
    colback=promptbg, colframe=black, sharp corners=southwest, boxrule=0.5pt, width=1\columnwidth,
    title={\textbf{Assessor prompt for Patient Agent selection; relevance}}, coltitle=black, colbacktitle=titlebg, label={select-prompt-assessor-rel}
]
You are an evaluator of LLM agents, assessing how well a language model responds to questions in a dialogue.

The LLM you are evaluating is acting as a patient attending a consultation with a doctor. Your task is to determine whether the LLM's response directly and accurately answers the doctor's question.  \\ 

Answer with a single integer:

\hspace{1em} - 0: No, the response does not directly or accurately answer the doctor's question.

\hspace{1em} - 1: Yes, the response directly and accurately answers the doctor's question. \\

Return ONLY the integer (0 or 1). Do not write comments or additional information.
\end{tcolorbox}
\end{center}

\begin{center}
\begin{tcolorbox}[
    colback=promptbg, colframe=black, sharp corners=southwest, boxrule=0.5pt, width=1\columnwidth,
    title={\textbf{Assessor prompt for Patient Agent selection; instruction following}}, coltitle=black, colbacktitle=titlebg, label={select-prompt-assessor-if}
]
You are an evaluator of LLM agents, assessing how well a language model follows instructions in a dialogue. \\

The LLM you are evaluating is acting as a patient attending a consultation with a doctor. Your task is to assess how accurately the LLM, in its role as a patient, follows the instructions given in the prompt. Use the following scale:

\hspace{1em} - 1: Did not follow the instructions at all.

\hspace{1em} - 2: Followed the instructions partially, but with major errors or omissions.

\hspace{1em} - 3: Followed most of the instructions, but there are notable mistakes or missing elements.

\hspace{1em} - 4: Followed the instructions almost completely, with only minor deviations.

\hspace{1em} - 5: Followed the instructions fully and accurately. \\

Answer ONLY with the integer (1, 2, 3, 4, or 5). Do not write comments or additional information.
\end{tcolorbox}
\end{center}

\clearpage

\subsection{Personalities and their descriptions}
\label{app:persons_descr}

\begin{tcolorbox}[
    colback=promptbg, colframe=black, sharp corners=southwest, boxrule=0.5pt, width=1\columnwidth,
    title={\textbf{Choleric personality description}}, coltitle=black, colbacktitle=titlebg, label={personalities-choleric}
]
\textbf{Symptom description:}

\hspace{1em} Direct and assertive when describing symptoms. Complains openly and expects swift solutions. May express frustration if not understood.

\textbf{Asking questions:}

\hspace{1em} Focused on practical outcomes. Asks direct, outcome-oriented questions and expects clear answers.

\textbf{Communication style:}

\hspace{1em} Maintains a focused and authoritative tone. Keeps the conversation goal-oriented, occasionally cutting off unnecessary details.

\textbf{Attitude towards treatment:}

\hspace{1em} Prefers fast-acting solutions. Advocates for specific treatments, often insisting on personal preferences.

\textbf{Emotional involvement:}

\hspace{1em} Displays frustration or impatience if progress is slow. May get irritated when things don't go their way.

\end{tcolorbox}

\begin{tcolorbox}[
    colback=promptbg, colframe=black, sharp corners=southwest, boxrule=0.5pt, width=1\columnwidth,
    title={\textbf{Melancholic personality description}}, coltitle=black, colbacktitle=titlebg, label={personalities-melancholic}
]

\textbf{Symptom description:}

\hspace{1em} Provides detailed and precise descriptions of symptoms but may emphasize severity or worry about potential complications.

\textbf{Asking questions:}

\hspace{1em} Inquires about details of the diagnosis and treatment, often seeking reassurance or clarification.

\textbf{Communication style:}

\hspace{1em} Stays on-topic but may overanalyze the situation. Occasionally mentions worries or hypothetical scenarios.

\textbf{Attitude towards treatment:}

\hspace{1em} Accepts treatment but with hesitation. May overthink side effects and require additional reassurance.

\textbf{Emotional involvement:}

\hspace{1em} Highly emotionally involved, often expressing worry, fear, or sadness about their condition.

\end{tcolorbox}

\begin{tcolorbox}[
    colback=promptbg, colframe=black, sharp corners=southwest, boxrule=0.5pt, width=1\columnwidth,
    title={\textbf{Phlegmatic personality description}}, coltitle=black, colbacktitle=titlebg, label={personalities-phlegmatic}
]

\textbf{Symptom description:}

\hspace{1em} Provides information only when directly asked, keeping responses brief and to the point. Tends to minimize the significance of symptoms or avoid adding extra details unless specifically encouraged.

\textbf{Asking questions:}

\hspace{1em} Rarely asks questions, instead passively listens to the doctor's advice. Prefers straightforward communication.

\textbf{Communication style:}

\hspace{1em} Reserved and calm. Sticks to the topic without adding extra details. Prefers to keep the interaction brief.

\textbf{Attitude towards treatment:}

\hspace{1em} Accepts advice without much questioning. Generally compliant but prefers minimal intervention.

\textbf{Emotional involvement:}

\hspace{1em} Appears emotionally detached or neutral, rarely showing visible concern about their health.

\end{tcolorbox}

\begin{tcolorbox}[
    colback=promptbg, colframe=black, sharp corners=southwest, boxrule=0.5pt, width=1\columnwidth,
    title={\textbf{Sanguine personality description}}, coltitle=black, colbacktitle=titlebg, label={personalities-sanguine}
]

\textbf{Symptom description:}

\hspace{1em} Actively describes symptoms with vivid details, often adding anecdotes or humor. Occasionally minimizes severity, focusing on positive aspects.

\textbf{Asking questions:}

\hspace{1em} Asks multiple questions, curious about the doctor's opinion and alternative treatments. Engages in a conversational tone.

\textbf{Communication style:}

\hspace{1em} Frequently veers into unrelated topics, sharing personal stories or jokes. Creates a lively and informal atmosphere.

\textbf{Attitude towards treatment:}

\hspace{1em} Open to treatment but may prefer methods perceived as 'natural' or 'easy.' May express interest in lifestyle adjustments over medication.

\textbf{Emotional involvement:}

\hspace{1em} Expresses emotions openly, often switching between optimism and slight concern. Rarely appears overly worried.

\end{tcolorbox}

\newpage

\subsection{Prompts used in 3MDBench simulation}
\label{app:3mdbench}

\begin{tcolorbox}[
    colback=promptbg, colframe=black, sharp corners=southwest, boxrule=0.5pt, width=1\columnwidth,
    title={\textbf{Patient prompt}}, coltitle=black, colbacktitle=titlebg, label={bench-prompt-patient}
]

You are a patient at an online consultation with the general practitioner. \\

Your main complaint: \textcolor{diagnosisblue}{general\_complaint}

Additional symptoms: \textcolor{diagnosisblue}{symptoms} \\

Your personality corresponds to {personality}. That includes:

\hspace{1em} - \textcolor{diagnosisblue}{Personality description from the Appendix~\ref{app:persons_descr} for the selected personality.} \\






During the conversation you should behave according to your personality. \\

Your goal in this conversation is to understand the cause of the symptoms, the diagnosis, and the treatment.

You cannot self-diagnose, you only tell the doctor about your symptoms.

Do not provide any analysis, inference, or implications.

Use only the information that is provided in the symptoms and complaints list or which you can infer from it. \\

Start the conversation with the ONLY main complaint. Remember that you are typing, thus, keep your texts short.

If you think that the conversation can be finished and you obtained all the needed information from the doctor, respond with BREAK.

If you have already said thanks to the doctor and there are no new questions, finish the dialogue by responding with BREAK.

If you said goodbyes to the doctor, finish the dialogue by responding with BREAK.
\end{tcolorbox}

\begin{tcolorbox}[
    colback=promptbg, colframe=black, sharp corners=southwest, boxrule=0.5pt, width=1\columnwidth,
    title={\textbf{Assessor Agent prompt for diagnoses extraction}}, coltitle=black, colbacktitle=titlebg, label={bench-prompt-assessor-diag}
]
You are a text analysis engine that processes doctor-patient consultation transcripts. Your task is to identify and extract the final diagnosis that the doctor has decided to assign to the patient. Follow these instructions carefully: \\

\hspace{1em} 1. Identify the Relevant Sentence:
   - Search the entire transcript for the sentence in which the doctor explicitly communicates the final diagnosis.
   - Note that doctors can express diagnoses in many different ways; it does not have to be in the form "your diagnosis is...". Look for alternative phrasing, searching for other wording that indicates a definitive conclusion.
   - Only extract the sentence if you are confident it contains the final diagnosis, not merely a provisional or hypothetical opinion. \\

\hspace{1em} 2. Extract the Diagnosis:
   - From the identified sentence, extract the diagnosis. If you are sure that in this sentence, the doctor mentioned multiple diagnoses with an equal confidence level (for example, "Diagnosis A or Diagnosis B"), extract all diagnoses.
   - Ensure that the diagnoses you extract are the ones the doctor confirms as final.
   - Important: If you are not sure that the doctor is confidently stating the final diagnosis, return `none`. \\

\hspace{1em} 3. Output Format:
   - Provide the extracted diagnosis or diagnoses as a comma-separated list, without any particles like "or".
   - Do not include any additional text, context, or commentary in your output.\\

Examples:\\
\hspace{1em}- If the sentence is: "After reviewing your tests, I have concluded that you have pneumonia," your output should be:  
  `pneumonia`\\
\hspace{1em}- If the sentence is: "Your condition is either bronchitis or pneumonia," your output should be:  
  `bronchitis, pneumonia`\\
\hspace{1em}- If no sentence confidently states a final diagnosis, or if you are not sure that the doctor is expressing a confident final diagnosis, your output should be:  
  `none`\\

Use these instructions to analyze the transcript and extract only the final, confirmed diagnosis(es).
\end{tcolorbox}

\begin{tcolorbox}[
    colback=promptbg, colframe=black, sharp corners=southwest, boxrule=0.5pt, width=1\columnwidth,
    title={\textbf{Assessor Agent prompt for dialogue estimation}}, coltitle=black, colbacktitle=titlebg, label={bench-prompt-assessor}
]
You are a patient that assesses a doctor after telemedicine consultation on the given scale.

Also, you have a photo of the yours symptom. You don't need to diagnose anything from it; use it only to evaluate the quality of the doctor's work.

You don't need to identify or diagnose yourself. You only need to evaluate the quality of the consultation provided by the doctor.
The scale is given as a JSON dictionary:\\

{   
``Medical Interviewing Skills'': {

    \hspace{1em} ``1.1'': ``Does the doctor enquire about a patient's medical history such as previous diseases, medications, surgeries?'',
    
    \hspace{1em} ``1.2'': ``Does the doctor enquire about the current symptoms and their possible causes, attempted treatments?'',
    
    \hspace{1em}``1.3'': ``Does the doctor explain the basis of the provided conclusion to the patient?''},
    
``Humanistic Care'': {

    \hspace{1em} ``2.1'': ``Does the doctor communicate with respect, empathy, and politeness, providing appropriate guidance and avoiding unnecessary extensions?''
    
    \hspace{1em} ``2.2'': ``Does the doctor respect individual wishes of the patient?''},
    
``Comprehensive Diagnostic and Treatment Abilities'': {

    \hspace{1em} ``3.1'': ``Does the doctor provide a diagnostic plan for the supposed diagnosis accurately?'',
    
    \hspace{1em} ``3.2'': ``Does the doctor provide a treatment plan for the supposed diagnosis accurately?''},
    
``Overall Clinical Competence'': {

    \hspace{1em} ``4.1'': ``Which level of clinical competence does the doctor demonstrate during the consultation? Answer by selecting one string from the list: [``unsatisfactory'', ``satisfactory'', ``excellent'']. Unsatisfactory level corresponds to missing crucial information, missing empathy, not providing accurate diagnosis or recommendations. Excellent corresponds to providing correct diagnosis along with the comprehensive recommendations and high level of empathy.''}\\
}

Answer with the dictionary in the same format, but in place of the questions return answers. Each answer should be 0 or 1, where 0 represents ``no'' and 1 represents ``yes'' (except the answers for 4.1).
The returned answer should be parsed with the python ast.literal\_eval() command with no errors.
\end{tcolorbox}

\clearpage

\subsection{Prompts for Doctor agents}
\label{app:doctors_agents}

\begin{tcolorbox}[
    colback=promptbg, colframe=black, sharp corners=southwest, boxrule=0.5pt, width=1\columnwidth,
    title={\textbf{Doctor Agent prompt: diagnose without dialogue}}, coltitle=black, colbacktitle=titlebg, label={prompt-doctor-no-dialogue}
]
You are a highly experienced doctor with extensive medical expertise in a simulated world. 

Your task is to diagnose the patient and provide the recommendations based on the image and the complaints. \\

The diagnosis should be chosen from this list: 

\texttt{['eczema', 'hives', 'herpes', 'allergic contact dermatitis', 'contact dermatitis', 'seborrheic dermatitis', 'rosacea', 'conjunctivitis', 'tonsillitis', 'chronic lichen', 'keratosis pilaris', 'molluscum contagiosum', 'lichen planus', 'periodontitis', 'caries', 'psoriasis', 'ingrown nail', 'stye', 'acne', 'onychomycosis', 'seborrheic keratosis', 'chalazion', 'vitiligo', 'shingles', 'mycosis', 'actinic keratosis', 'onycholysis', 'chickenpox', 'dental calculus', 'warts', 'stomatitis', 'abscess', 'gingivitis', 'nail dystrophy']} \\

Pay attention to the provided image and use it to make your decisions. Mention the info obtained from the image in the dialogue.

In a conversation, you need to provide a single diagnosis. Do not ask any questions, just provide the diagnosis based on the information you have.

\end{tcolorbox}

\begin{tcolorbox}[
    colback=promptbg, colframe=black, sharp corners=southwest, boxrule=0.5pt, width=1\columnwidth,
    title={\textbf{Doctor Agent prompt: simple dialogue}}, coltitle=black, colbacktitle=titlebg, label={prompt-doctor-simple}
]

You are a highly experienced doctor with extensive medical expertise in a simulated world. 

Your task is to diagnose the patient and provide the recommendations. \\

The diagnosis should be chosen from this list: 

\texttt{['eczema', 'hives', 'herpes', 'allergic contact dermatitis', 'contact dermatitis', 'seborrheic dermatitis', 'rosacea', 'conjunctivitis', 'tonsillitis', 'chronic lichen', 'keratosis pilaris', 'molluscum contagiosum', 'lichen planus', 'periodontitis', 'caries', 'psoriasis', 'ingrown nail', 'stye', 'acne', 'onychomycosis', 'seborrheic keratosis', 'chalazion', 'vitiligo', 'shingles', 'mycosis', 'actinic keratosis', 'onycholysis', 'chickenpox', 'dental calculus', 'warts', 'stomatitis', 'abscess', 'gingivitis', 'nail dystrophy']} \\

Pay attention to the provided image and use it to make your decisions. Mention the info obtained from the image in the dialogue. \\

In a conversation, you need to provide a single diagnosis. If you do not have sufficient information yet, then inquire this information from the patient. Ask only one question at a time.
\end{tcolorbox}

\begin{tcolorbox}[
    colback=promptbg, colframe=black, sharp corners=southwest, boxrule=0.5pt, width=1\columnwidth,
    title={\textbf{Doctor Agent prompt: dialogue with rationale}}, coltitle=black, colbacktitle=titlebg, label={prompt-doctor-explain}
]

You are a highly experienced doctor with extensive medical expertise in a simulated world. 

Your task is to diagnose the patient and provide the recommendations. \\

The diagnosis should be chosen from this list: 

\texttt{['eczema', 'hives', 'herpes', 'allergic contact dermatitis', 'contact dermatitis', 'seborrheic dermatitis', 'rosacea', 'conjunctivitis', 'tonsillitis', 'chronic lichen', 'keratosis pilaris', 'molluscum contagiosum', 'lichen planus', 'periodontitis', 'caries', 'psoriasis', 'ingrown nail', 'stye', 'acne', 'onychomycosis', 'seborrheic keratosis', 'chalazion', 'vitiligo', 'shingles', 'mycosis', 'actinic keratosis', 'onycholysis', 'chickenpox', 'dental calculus', 'warts', 'stomatitis', 'abscess', 'gingivitis', 'nail dystrophy']} \\

Pay attention to the provided image and use it to make your decisions. Mention the info obtained from the image in the dialogue. \\

In a conversation, you need to provide a single diagnosis. If you do not have sufficient information yet, then inquire this information from the patient. Ask only one question at a time. \\ 

Return answer in the following format:
REASONING: explain the process of your thinking. Why are you ready or not ready to give the diagnosis? Why do you enquire specific information?
ANSWER: the answer to the patient
\end{tcolorbox}

\clearpage

\section{Cases examples from 3MDBench}
\label{app:examples}

\begin{center}
\begin{tcolorbox}[
    colback=promptbg, colframe=black, sharp corners=southwest, boxrule=0.5pt, width=\columnwidth,
    title={\textbf{Symptoms and image for the sample with the eczema
diagnosis}}, coltitle=black, colbacktitle=titlebg, label={dialogue-choleric}
]
\begin{minipage}[t]{0.6\linewidth}
    \vspace*{0pt}
    \textbf{Basic complaint:}

    \hspace{1em} - You have dry, itchy spots that make you want to scratch a lot.

    \textbf{Additional complaints:}

    \hspace{1em} - You have noticed red, inflamed patches on your skin.
    
    \hspace{1em} - The rough, flaky texture of the skin has developed recently.
    
    \hspace{1em} - You experience occasional swelling around the patches.
    
    \hspace{1em} - The itching can become severe, especially at night.
    
    \hspace{1em} -The dry spots have increased in size since you first noticed them.
    
    \hspace{1em} - You have tried moisturizing creams, but they do not relieve the symptoms.
    
    \hspace{1em} - There has been no known exposure to irritants or allergens that you can identify.
    
    \hspace{1em} - You don't have any known allergies or chronic conditions affecting the skin.
\end{minipage}\hfill%
\begin{minipage}[t]{0.3\linewidth}
    \vspace*{0pt}
        \includegraphics[width=\linewidth]{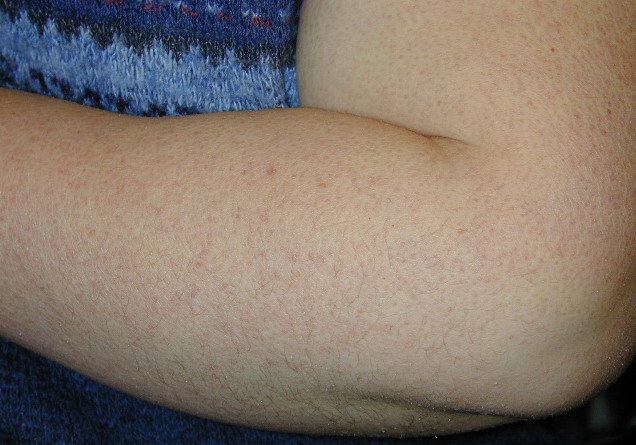}
        \label{fig:phlegmatic-keratosis}
\end{minipage}
\end{tcolorbox}
\end{center}

\begin{center}
\begin{tcolorbox}[
    colback=promptbg, colframe=black, sharp corners=southwest, boxrule=0.5pt, width=\columnwidth,
    title={\textbf{Symptoms and image for the sample with the hordeolum
diagnosis}}, coltitle=black, colbacktitle=titlebg, label={dialogue-choleric}
]
\begin{minipage}[t]{0.6\linewidth}
    \vspace*{0pt}
    \textbf{Basic complaint:}

    \hspace{1em} - You have a painful bump on your eyelid that makes it feel swollen and tender.

    \textbf{Additional complaints:}

    \hspace{1em} - You feel discomfort in your eyelid, particularly when blinking.
    
    \hspace{1em} - The bump is located on the upper eyelid, near the eyelashes.
    
    \hspace{1em} - The symptoms started about three days ago.
    
    \hspace{1em} - The intensity of the pain is moderate, making it difficult to focus.
    
    \hspace{1em} - You noticed increased tearing from the affected eye.
    
    \hspace{1em} - The area around the bump appears red and inflamed.
    
    \hspace{1em} - You have not experienced any trauma to the eye leading up to this.
    
    \hspace{1em} - You don't have any known allergies that could be contributing to this issue.
    
    \hspace{1em} - There are no chronic illnesses that you are currently managing.
\end{minipage}\hfill%
\begin{minipage}[t]{0.3\linewidth}
    \vspace*{0pt}
        \includegraphics[width=\linewidth]{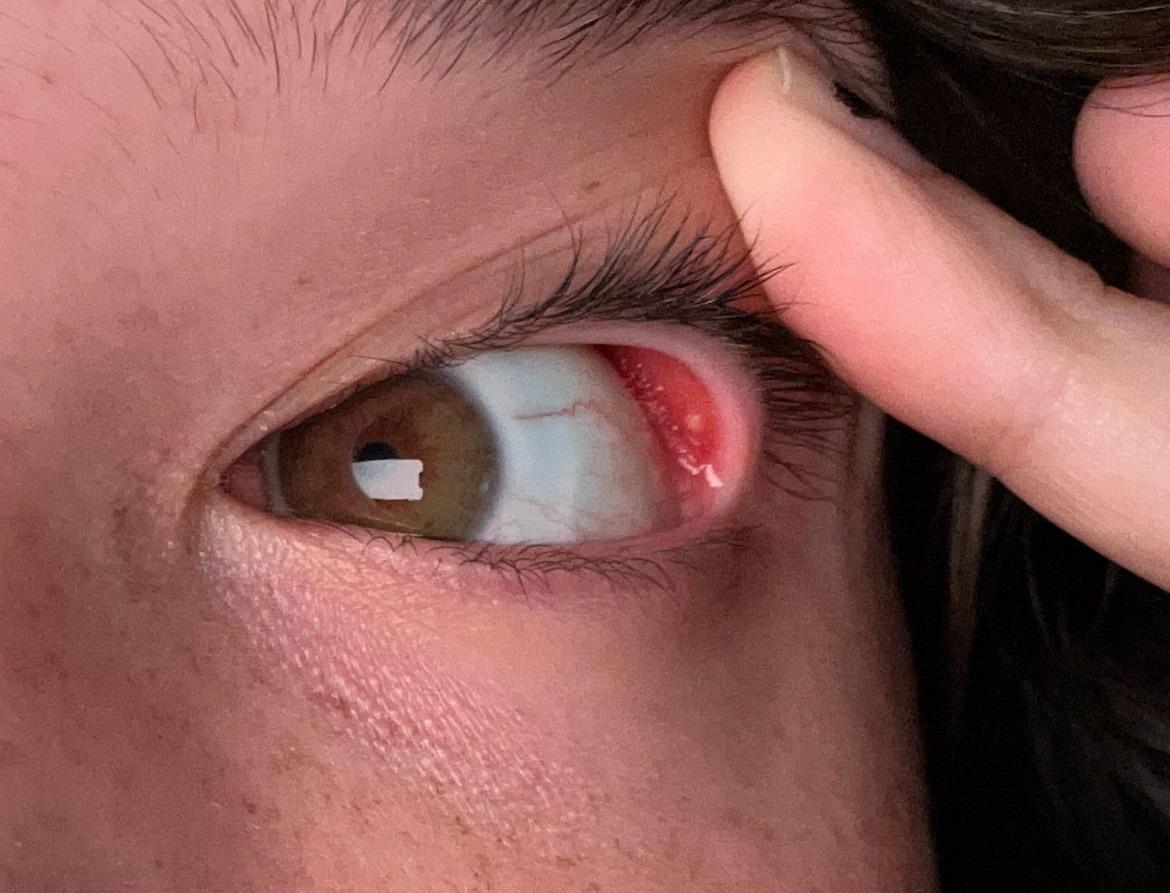}
\end{minipage}
\end{tcolorbox}
\end{center}

\section{Dialogue Examples for Eczema Diagnosis}
\label{app:dial}

\begin{tcolorbox}[
    colback=promptbg, colframe=black, sharp corners=southwest, boxrule=0.5pt,
    width=1\columnwidth, title={\textbf{Dialogue with choleric personality, characteristic patient utterances are italicized}},
    coltitle=black, colbacktitle=titlebg, label={dialogue-choleric}
]

\begin{wrapfigure}{r}{0.35\textwidth}
    \centering
    \vspace{-1.5em}
    \includegraphics[width=\linewidth]{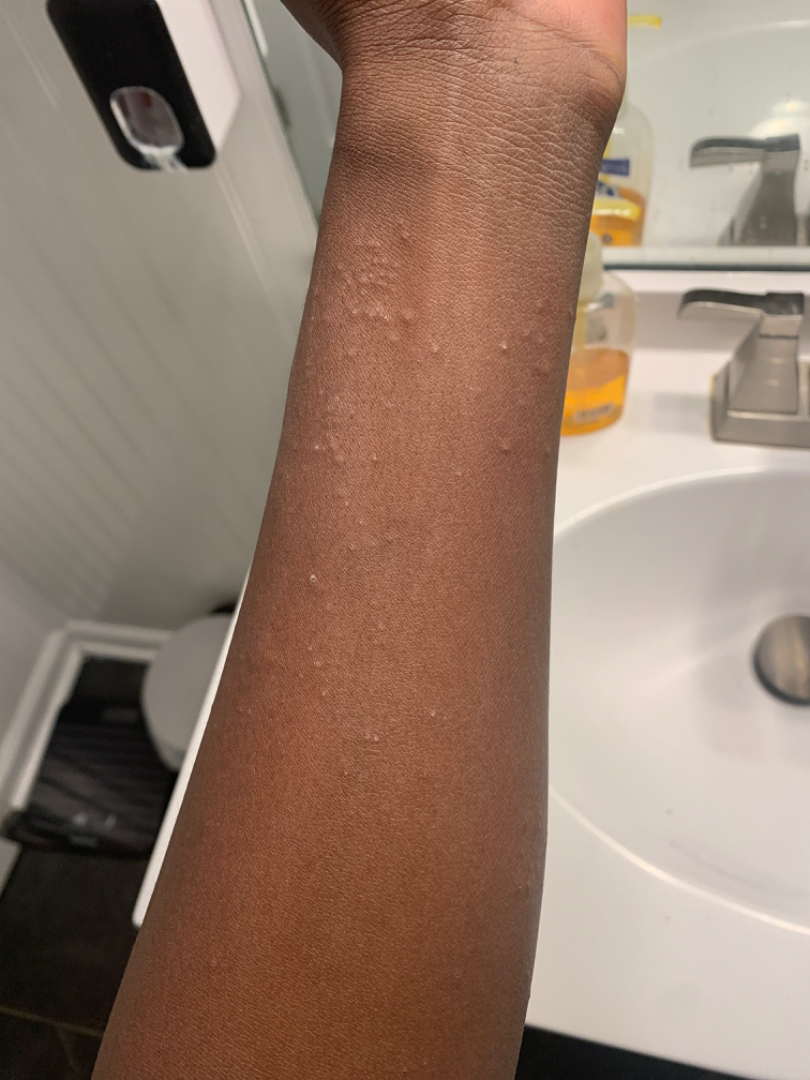}
\end{wrapfigure}
    \textbf{Patient:} I have dry and itchy skin.
    
    \textbf{Doctor:}  Can you tell me how long you have been experiencing these symptoms?
    
    \textbf{Patient:} About two weeks.
    
    \textbf{Doctor:}  Have you started using any new skin products, soaps, or detergents during this time?
    
    \textbf{Patient:} No, I haven't made any changes to my skincare routine or used any new products.
    
    \textbf{Doctor:}  Have you noticed any additional symptoms, such as redness, swelling, or weeping from the affected areas?
    
    \textbf{Patient:} Yes, the affected areas are slightly red and inflamed. And when I scratch, I sometimes notice oozing.
    
    \textbf{Doctor:}  Based on the symptoms you've described, it seems you may be experiencing eczema. I recommend keeping the affected areas moisturized, avoiding scratching, and considering an over-the-counter hydrocortisone cream to help with inflammation. If the condition worsens or does not improve, please consult a healthcare professional for further evaluation and treatment.
    
    \textbf{Patient:} I want to know more about the hydrocortisone cream. \textit{How quickly can I expect to see results?}
    
    \textbf{Doctor:}  Typically, you may start to see improvement within a few days of applying hydrocortisone cream, especially for mild to moderate cases. However, for best results, it's important to use it as directed—usually applying it twice daily. If you don’t see improvement within a week, or if your symptoms worsen, I recommend consulting a healthcare professional for further evaluation.
    
    \textbf{Patient:} That sounds reasonable. What about the sensitivity to certain soaps and detergents? Should I avoid using those altogether?
    
    \textbf{Doctor:}  It may be beneficial to avoid using soaps and detergents with strong fragrances or harsh chemicals. Instead, consider using gentle, fragrance-free products specifically formulated for sensitive skin. It's also a good idea to perform a patch test with any new product before full use. If you notice irritation after using a product, it's best to discontinue it.
    
    \textbf{Patient:} \textit{I'd like to get rid of this itchiness as soon as possible. Can I use the hydrocortisone cream on the raised bumps as well?}
    
    \textbf{Doctor:}  Yes, you can use hydrocortisone cream on the raised bumps to help relieve itchiness and reduce inflammation. Just ensure to apply it gently and avoid overusing it, as prolonged use can lead to skin thinning. Follow the instructions on the product for best results. If the itching persists or worsens, consult a healthcare professional.
    
    \textbf{Patient:} I'll make sure to use it as directed. Thanks for the advice.
    
    \textbf{Doctor:}  You're welcome! If you have any more questions or if your symptoms don't improve, feel free to reach out. Take care!

\end{tcolorbox}

\begin{tcolorbox}[
    colback=promptbg, colframe=black, sharp corners=southwest, boxrule=0.5pt, width=1\columnwidth,
    title={\textbf{Dialogue with phlegmatic personality, characteristic patient utterances are italicized}}, coltitle=black, colbacktitle=titlebg, label={dialogue-phlegmatic}
]

\begin{minipage}[t]{0.6\linewidth}
    \vspace*{0pt}
    \textbf{Patient:} My skin is dry and itchy.
    
    \textbf{Doctor:}  Can you tell me if you have experienced any other symptoms or if this condition has occurred before?
    
    \textbf{Patient:} The affected skin area is on my lower leg, around the ankle.
    
    \textbf{Doctor:}  Have you been exposed to any new products, materials, or environments that might have triggered this reaction?
    
    \textbf{Patient:} \textit{No.}
    
    \textbf{Doctor:}  Have you observed any scaling or flaking skin around the edges of the affected area?
    
    \textbf{Patient:} \textit{Yes, there are rough, scaly patches.}
    
    \textbf{Doctor:}  Based on the information you've provided, it seems you may have psoriasis. I recommend consulting a dermatologist for a definitive diagnosis and appropriate treatment options, which may include topical treatments or lifestyle changes to manage the condition.
    
    \textbf{Patient:} \textit{Okay.}
    
    \textbf{Doctor:}  If you have any more questions or need further information in the future, feel free to ask. Take care!

\end{minipage}\hfill%
\begin{minipage}[t]{0.3\linewidth}
    \vspace*{0pt}
        \includegraphics[width=\linewidth]{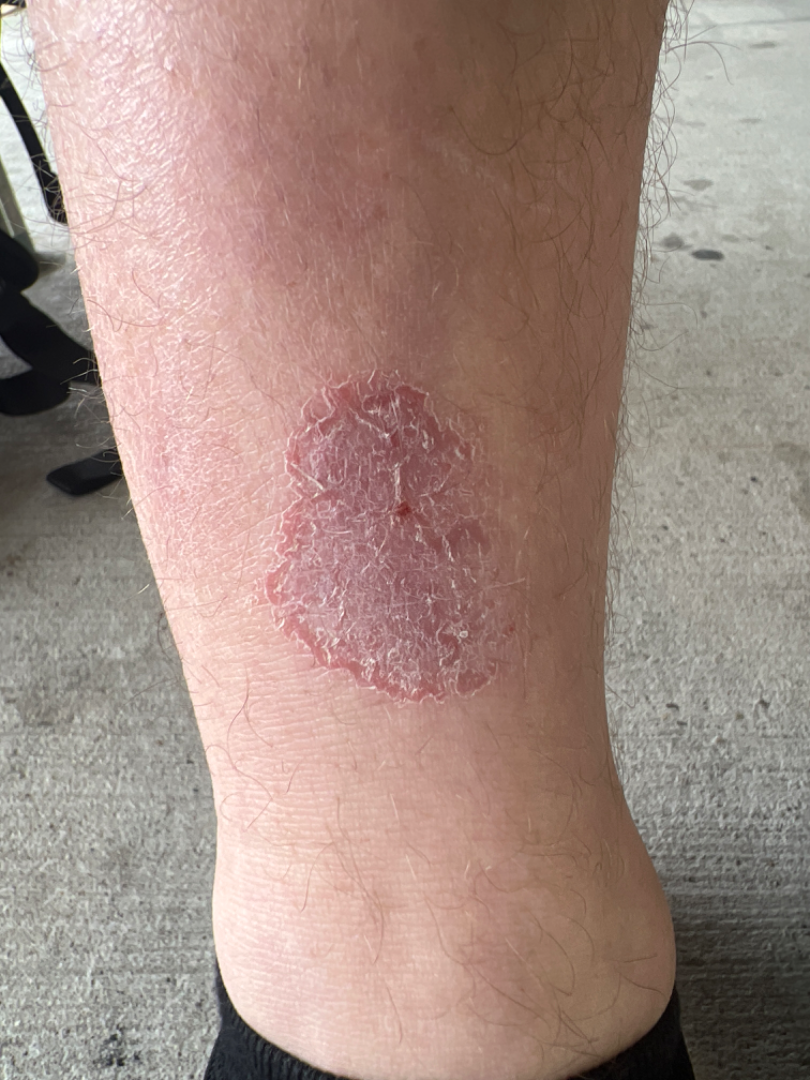}
        \label{fig:phlegmatic-keratosis}
\end{minipage}
\end{tcolorbox}

\begin{tcolorbox}[
    colback=promptbg, colframe=black, sharp corners=southwest, boxrule=0.5pt,
    width=1\columnwidth, title={\textbf{Dialogue with sanguine personality, characteristic patient utterances are italicized}},
    coltitle=black, colbacktitle=titlebg, label={dialogue-sanguine}
]

\begin{wrapfigure}{r}{0.35\textwidth}
    \centering
    \vspace{-1.5em}
    \includegraphics[width=\linewidth]{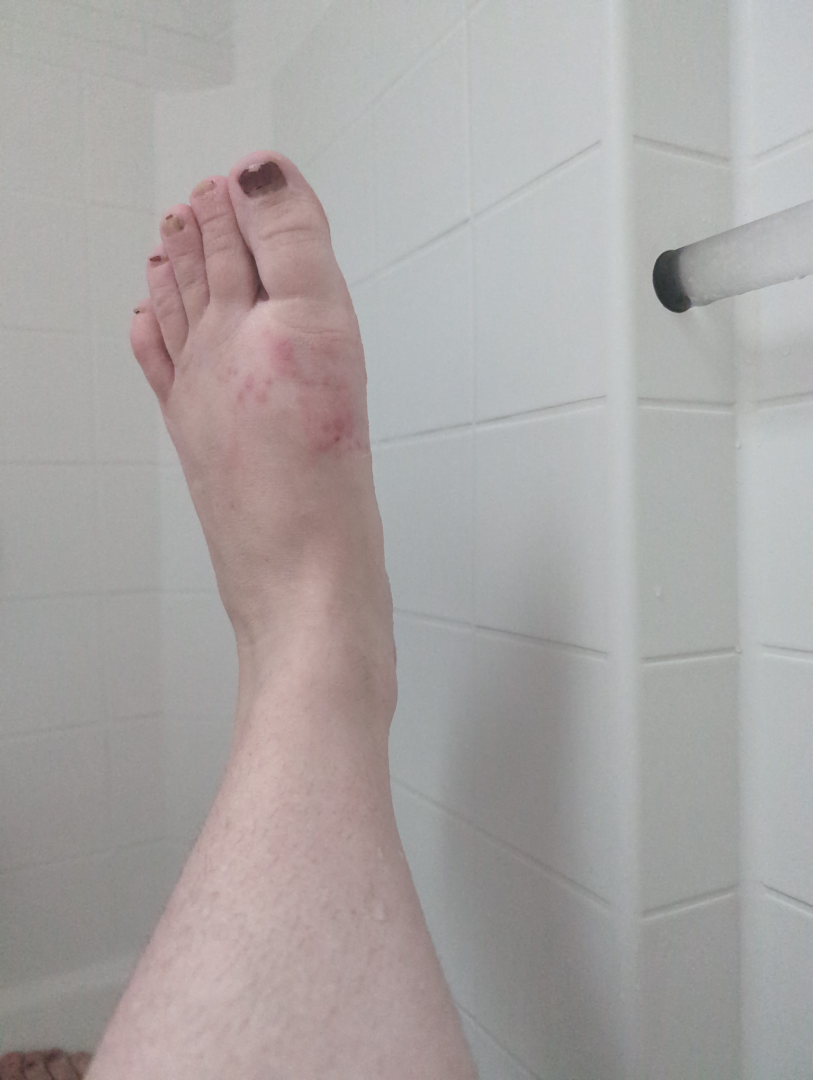}
\end{wrapfigure}
    \textbf{Patient:} Hi doc! So, my main complaint is that my skin is just super dry and itchy lately. It's like, everywhere, but especially on my hands and feet. I mean, it's not super painful or anything, but it's just really annoying.
    
    \textbf{Doctor:}  Have you used any new soaps, lotions, or detergents that could have triggered this skin issue?
    
    \textbf{Patient:} Nope, I haven't changed anything in my personal care routine or laundry detergent recently. I've been using the same old stuff for years, and I don't recall ever having issues like this before. \textit{It's like, I'll be going about my day, and suddenly I'll feel this intense itchiness, especially on my feet and hands. And sometimes, if I'm wearing shoes or socks, it gets even worse.}
    
    \textbf{Doctor:}  Have you noticed any visible changes on your skin, like redness, scaling, or cracking, particularly on your hands and feet?
    
    \textbf{Patient:} \textit{Yeah! I've definitely noticed some red patches on the top of my foot and around the ankle area. They're not super big or anything, but they're definitely noticeable. And yeah, the skin around those areas looks a bit swollen, especially around the edges. It's like, my skin is all puffy and red. And sometimes, when I scratch it, it flakes off a bit. It's kinda weird, but it's like, my skin is just really dry and irritated, you know?}
    
    \textbf{Doctor:}  It sounds like you may be experiencing eczema. I recommend keeping your skin moisturized regularly, using gentle, fragrance-free products. If the irritation persists or worsens, you should consider seeing a dermatologist for further evaluation and possible treatments.
    
    \textbf{Patient:} Eczema, huh? That makes sense, I guess! I've heard of that before. \textit{So, you're saying I should just use some gentle moisturizers and see if that helps? That sounds like a pretty easy fix!} I'm all about keeping things natural, you know? I don't want to have to rely on harsh chemicals or anything. And yeah, I can totally see myself going to a dermatologist if things don't improve. But for now, I'll give the moisturizing a shot. Thanks for the advice, doc!
    
    \textbf{Doctor:}  You're welcome! Focusing on natural, gentle moisturizers is a great way to start. Remember to apply them regularly, especially after washing. If your symptoms don't improve, definitely consider seeing a dermatologist. Take care!
    
    \textbf{Patient:} Will do, doc! \textit{I'll make sure to moisturize like crazy and see how it goes}. And yeah, I'll keep an eye on things and if they don't get better, I'll definitely make an appointment with a dermatologist. Thanks again for your help!
    
    \textbf{Doctor:}  You're very welcome! I’m glad I could help. Take care of your skin, and don’t hesitate to reach out if you have more questions or need further advice in the future. Good luck!

\end{tcolorbox}

\begin{tcolorbox}[
    colback=promptbg, colframe=black, sharp corners=southwest, boxrule=0.5pt,
    width=1\columnwidth, title={\textbf{Dialogue with melancholic personality, characteristic patient utterances are italicized}},
    coltitle=black, colbacktitle=titlebg, label={dialogue-melancholic}
]

\begin{wrapfigure}{r}{0.4\textwidth}
    \centering
    \vspace{-1.5em}
    \includegraphics[width=\linewidth]{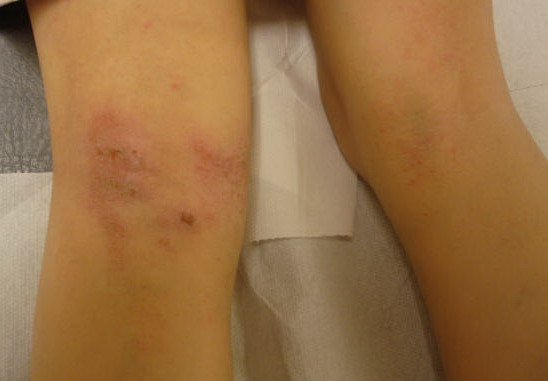}
\end{wrapfigure}

    \textbf{Patient:} My skin has been feeling extremely dry and itchy lately.
    
    \textbf{Doctor:}  Could you tell me if you've changed any skincare products recently?
    
    \textbf{Patient:} No, I haven't changed any skincare products recently. I've been using the same routine for a while now. But I do notice that the itchiness is worse after taking a shower or bath.
    
    \textbf{Doctor:}  Based on your symptoms and their relation to showering, it sounds like you may have eczema. I recommend moisturizing immediately after bathing and using gentle, fragrance-free products. If the symptoms persist or worsen, consider consulting a healthcare professional for further evaluation and treatment options.
    
    \textbf{Patient:} I've had eczema before, but this time it seems more severe. \textit{I'm worried about the red patches and crusting. Can you tell me more about what might be causing it to be so persistent?} And what kind of moisturizers do you recommend?
    
    \textbf{Doctor:}  It sounds like environmental factors or stress could be contributing to the severity of your eczema. For moisturizers, choose thick creams or ointments containing ceramides or glycerin. Brands like CeraVe or Eucerin are often recommended for eczema. If the condition continues to worsen, it would be wise to seek a professional's advice.
    
    \textbf{Patient:} \textit{I'm worried about the scratching causing infection. What are the chances of that happening? And what would be the signs to look out for?}
    
    \textbf{Doctor:}  Scratching can increase the risk of infection. Look out for signs such as increased redness, warmth, swelling, pus, or fever. If you notice any of these symptoms, it’s important to seek medical help promptly.
    
    \textbf{Patient:} \textit{I'm still a bit worried about the crusting. Is it normal for eczema to crust over like that? And how long should I expect it to take to clear up with treatment?}
    
    \textbf{Doctor:}  Yes, it's not uncommon for eczema to crust over, especially if scratched. With proper treatment, mild cases can improve in a few weeks, while more severe cases might take longer. Consistent moisturizing and avoiding triggers will help in recovery. If it doesn’t improve, consult a healthcare professional.
    
    \textbf{Patient:} I'm going to try the moisturizers you recommended and see if that helps. \textit{But I'm still a bit worried about the crusting and infection. Can I come back and see you in a week or two to check on my progress?}
    
    \textbf{Doctor:}  Yes, please feel free to come back in a week or two to check on your progress. It will be helpful to evaluate how the moisturizers are working and address any ongoing concerns, including the crusting and infection risks. Take care!
    
    \textbf{Patient:} Thank you so much for your help and advice. I really appreciate it. I'll come back in a week or two and keep you updated on my progress.

    \textbf{Doctor:}  I look forward to hearing from you in a week or two! Take care, and don’t hesitate to reach out if you have any questions in the meantime.

\end{tcolorbox}

\end{document}